\documentclass[%
 reprint,
 amsmath,amssymb,
 aps,
]{revtex4-2}

\usepackage{graphicx}
\usepackage{dcolumn}
\usepackage{bm}

\usepackage{amsmath}
\usepackage{amssymb}
\usepackage{ragged2e}
\usepackage{physics}
\usepackage{soul}
\justifying

\usepackage{xcolor}
\usepackage{braket}
\usepackage{subfigure}
\usepackage{threeparttable}
\usepackage{amsthm}
\usepackage{float}
\usepackage[dvipsnames]{xcolor}
\usepackage{tikz,lipsum,lmodern}
\usepackage[most]{tcolorbox}
\usepackage{mathrsfs}
\usepackage[hidelinks]{hyperref}

\begin{document}

\preprint{APS/123-QED}

\title{Analytical solution of boundary time crystals via the superspin basis}

\author{Dominik Nemeth} 
\email{dominik.nemeth@manchester.ac.uk}
\author{Alessandro Principi}
\author{Ahsan Nazir}
\affiliation{%
Department of Physics and Astronomy,\\
\protect \mbox{The University of Manchester, Oxford Road, Manchester M13 9PL, United Kingdom}
}%

\date{\today}

\begin{abstract}
Boundary time crystals (BTCs) in dissipative collective spin systems have been extensively studied using numerical, mean-field, and perturbative approaches. However, an explicit Liouvillian description governing the long-time dynamics deep within the time crystal phase has remained elusive. Here, we derive an effective Liouvillian that analytically captures the extreme BTC regime, where dissipation is parametrically weak and oscillatory order is maximally robust. By introducing a superspin representation of Liouville space, we obtain closed-form expressions for the Liouvillian eigenvalues to first order in the dissipation strength, providing direct access to decay rates, oscillation frequencies, and their thermodynamic scaling. Applying this framework to the canonical BTC model we analytically recover spontaneous breaking of continuous time-translation symmetry and persistent oscillations in the thermodynamic limit. In contrast, we show that other dissipative spin models exhibit only single-frequency oscillatory dynamics and therefore do not support genuine BTC phases. Our results establish a controlled analytical framework for the long-time dynamics in the extreme BTC regime.
\end{abstract}

\maketitle

\section{Introduction}
\label{sec:introduction}
The study of time crystals arises from the question of whether spontaneous symmetry breaking—a key concept in condensed matter physics \cite{SSB1,SSB2,SSB3}—can be extended to the time domain. Spatial crystals break translational symmetry by forming periodic structures in space. Time crystals, on the other hand, break time-translation symmetry, either continuously or discretely, by exhibiting persistent oscillations in their steady state \cite{khemani2019briefhistorytimecrystals, Sacha_2018}. This concept was first proposed by Wilczek in 2012 \cite{wilczek} and was followed by significant works discussing the possibility of time-translation symmetry breaking in equilibrium and non-equilibrium systems \cite{significant_debate_1, significant_debate_2, significant_debate_3, significant_debate_4, no_go_theorem, Long_range_time_crystals}. This led to the discovery of discrete time crystals (DTCs), which emerge in periodically driven, out-of-equilibrium systems and exhibit robust subharmonic oscillations—oscillations at a period that is an integer multiple of the driving period \cite{DTC1,DTC2,DTC3,DTC4,DTC5,DTC6,DTC7,DTC8,DTC9, DTC10}. Experimental observations of these oscillations have been made in trapped ions, superconducting qubits, and many other systems \cite{DTC_EXPT_1,DTC_EXPT_2,DTC_EXPT_3,DTC_EXPT_4,DTC_EXPT_5}.

Building on these developments, the concept of time crystals has been extended to dissipative quantum systems. This new class of time crystals, known as boundary time crystals (BTCs), emerges in open quantum systems, in contrast to traditional time crystals realised in isolated settings \cite{BTC1, BTCs_d_level_system}. In BTCs, the interplay between coherent unitary dynamics and environmental dissipation can give rise to spontaneous breaking of continuous time-translation symmetry. As a consequence, the system exhibits persistent oscillations characterised by an emergent discrete time-translation symmetry, whose period is set by the least-damped eigenmode. The theoretical description of BTCs naturally relies on the Lindblad master equation, which governs the dynamics of open quantum systems.

Central to this phenomenon is the Liouvillian superoperator, which generates the system’s evolution in Liouville space. Its eigenvalue spectrum determines the long-time behaviour: the real parts of the eigenvalues control relaxation rates, while the imaginary parts generate oscillatory dynamics. In the BTC phase, the thermodynamic limit is accompanied by the emergence of purely imaginary Liouvillian eigenvalues, indicating persistent, non-decaying oscillations at long times. These spectral features signal the breaking of continuous time-translation symmetry and define the extreme time-crystal regime, where dissipation is parametrically weak and oscillatory order is maximally robust.

Most existing studies of BTCs have relied on numerical simulations and semiclassical approaches to identify the phase and characterise its properties \cite{franco_nori_btc,BTC_semiclassical_1,BTC_quantum_trajectories, exact_differential_method, btc_found, btc_numerical_1,btc_numerical_2, btc_numerical_3, btc_semiclassical_2, dissipative_dynamics_xxz_rg, btc_semiclassical_3, gapless_excitations_cumulant_expansion, btc_briefly_mentioned, wang2025boundarytimecrystalsinduced, dissipative_freezing, btc_quantum_thermo}. While mean-field and semiclassical descriptions provide valuable analytical insight into the dynamics of collective observables, they do not capture the microscopic Liouvillian spectrum. Analytical progress has nevertheless been achieved by treating dissipation perturbatively in the weak-coupling regime \cite{btcs_pt_nakanishi}, where it was claimed that exactly balanced gain and loss in the $x$ basis give rise to BTCs. However, this approach reduces the problem to tridiagonal effective matrices within degenerate Liouvillian subspaces, which generally do not admit closed-form expressions for their spectra. As a result, controlled analytical access to a Liouvillian description deep inside the BTC phase has remained limited.

In this work, we overcome this limitation by deriving an explicit effective Liouvillian that governs the long-time dynamics inside the BTC regime. By identifying a superspin representation of Liouville space, we directly diagonalise the perturbative Liouvillian and obtain closed-form expressions for its eigenvalues to first order in the dissipation strength. This provides controlled analytical access to decay rates, oscillation frequencies, and their thermodynamic scaling, and enables analytical calculations within the extreme BTC regime that were previously inaccessible.

Applying this framework to the canonical BTC model, we recover spontaneous breaking of continuous time-translation symmetry and undamped oscillations in the thermodynamic limit entirely analytically. In contrast, when applying the same effective Liouvillian approach to other dissipative spin models, we find that persistent oscillations alone do not guarantee genuine BTC behaviour. By solving the equations of motion for collective spin observables exactly, we show that models previously identified as BTCs (e.g. ``model B'' below) exhibit only single-frequency oscillatory dynamics, in sharp contrast to the multifrequency structure characteristic of true BTCs. In particular, we demonstrate that models B and C do not realise a BTC phase despite exhibiting gapless Liouvillian spectra.

This paper is organised as follows. In Sec.~\ref{sec:btc_model}, we introduce a widely studied collective spin model exhibiting BTC behaviour. In Sec.~\ref{sec:superspin_method}, we present the superspin representation and derive the effective Liouvillian governing the extreme BTC regime. In Sec.~\ref{sec:other_models}, we apply the same analytical framework to other dissipative spin Liouvillians and assess their long-time dynamics. We summarise our findings and discuss their implications in Sec.~\ref{sec:conclusions}. All analytical results are benchmarked against numerically exact master equation solutions obtained using QuTiP \cite{JOHANSSON20121760, JOHANSSON20131234, lambert2024qutip5quantumtoolbox}. Throughout this work, we use dimensionless units.

\section{BTC Model}
\label{sec:btc_model}
BTCs are usually discussed by considering the dynamics of a collective spin system interacting with an environment. The original idea was that the BTC lives on the boundary of a system whose bulk is the environment \cite{BTC1}. In the weak-coupling limit, the dynamics can be described by a Lindblad master equation, which in Ref.~\cite{BTC1} is of the form ($\hbar=1$)
\begin{equation} \label{eq:nori_model}
    \frac{d}{dt}\rho = \mathcal{L}\rho= -i[H_S, \rho] + N \Gamma(J_-\rho J_+ - \frac{1}{2}\{J_+ J_-, \rho\}).
\end{equation}
Here, we adopt the conventions employed in Ref.~\cite{franco_nori_btc}, such that for $\alpha=x,y,z$, $J_{\alpha}= (2/N )\sum_{i=1}^N \sigma_\alpha^{(i)}/2$ is a collective spin operator, $\sigma_{\alpha}^{(i)}$ is a Pauli matrix representing the $i^{\mathrm{th}}$ spin, $N$ is the number of spins, and the Liouvillian $\mathcal{L}$ governs the dynamics of the reduced density matrix of the spin system, $\rho$. The collective spin operators obey the commutation relations $[J_{\alpha}, J_{\beta}]= 2i \,\epsilon_{\alpha\beta\gamma} J_\gamma/N$ and $J_\pm = J_x \pm i J_y$. The system is described by the Hamiltonian $H_S = -N \Omega_x J_x$ and undergoes an incoherent process described by the Lindblad jump operator $J_-$ with a corresponding rate $\Gamma$. The collective spin operators describe $N$ spin-$1/2$ objects. We work in the maximally polarised subsector such that $J^2 =(2/N)^2 j(j+1)$ with $j=N/2$. The properties of these collective operators are discussed in more detail in \hyperref[appendix:collective_spin_operators]{Appendix} \ref{appendix:collective_spin_operators}.

\section{Perturbative Treatment in the Superspin Basis}
\label{sec:superspin_method}
A powerful approach to solving Lindblad master equations is the superoperator formalism. This makes use of an isomorphism between the linear space of operators defined on a $d$-dimensional Hilbert space, $\mathscr{H}_d$, and superoperators defined on the $d^2$-dimensional Liouville space, $\mathscr{L}_{d^2}$. For an arbitrary operator $\ket{m}\bra{n}$ defined on $\mathscr{H}_d$ we have an associated superket $\ket{m,n}\rangle = \ket{m}\otimes \ket{n}^*$ defined on $\mathscr{L}_{d^2}$. Following the transformation of $\rho$ to $\ket{\rho}\rangle$ outlined in \cite{Gyamfi_2020}, $\mathcal{L}$ is rewritten such that now it acts on $\ket{\rho}\rangle$. The eigenvalue equation for the superoperator $\mathcal{L}$ is then given by
\begin{equation}
   \mathcal{L}\ket{r_k}\rangle= (\mathcal{L}_0 + \mathcal{L}_D) \ket{r_k}\rangle = \lambda_k \ket{r_k}\rangle,
\end{equation}
where $\ket{r_k}\rangle$ is the right eigen-superket of the non-Hermitian $\mathcal{L}$ with eigenvalue $\lambda_k$ \cite{Kato:1966:PTL}, with
\begin{equation}
    \mathcal{L}_0= -i(H_S\otimes \mathbb{I} - \mathbb{I}\otimes H_S^{T} )
    \label{definition_L0}
\end{equation}
and
\begin{equation}
    \mathcal{L}_D = \sum_i \gamma_i \left(A_i \otimes A_i^* - \frac{1}{2}(A_i^{\dagger} A_i \otimes \mathbb{I} + \mathbb{I} \otimes (A_i^{\dagger} A_i)^T )\right).
    \label{genera_form_of_L_D}
\end{equation}
Here, $\mathbb{I}$ is the identity, $A_i$ denote generic Lindblad jump operators and $\gamma_i$ are their corresponding rates. Coherent dynamics is represented via the superoperator $\mathcal{L}_0$ and dissipative dynamics via $\mathcal{L}_D$. In the weakly dissipative regime, one can treat the dissipation as a perturbation to the coherent dynamics \cite{Perturbative_appraoch_open_quantum}. Our aim is to find first-order corrections to the eigenvalues of $\mathcal{L}_0$ by performing a power series expansion in $\gamma_i$, or in our case, $\Gamma$.

\begin{figure}[t]
    \centering
    \includegraphics[width=1\linewidth]{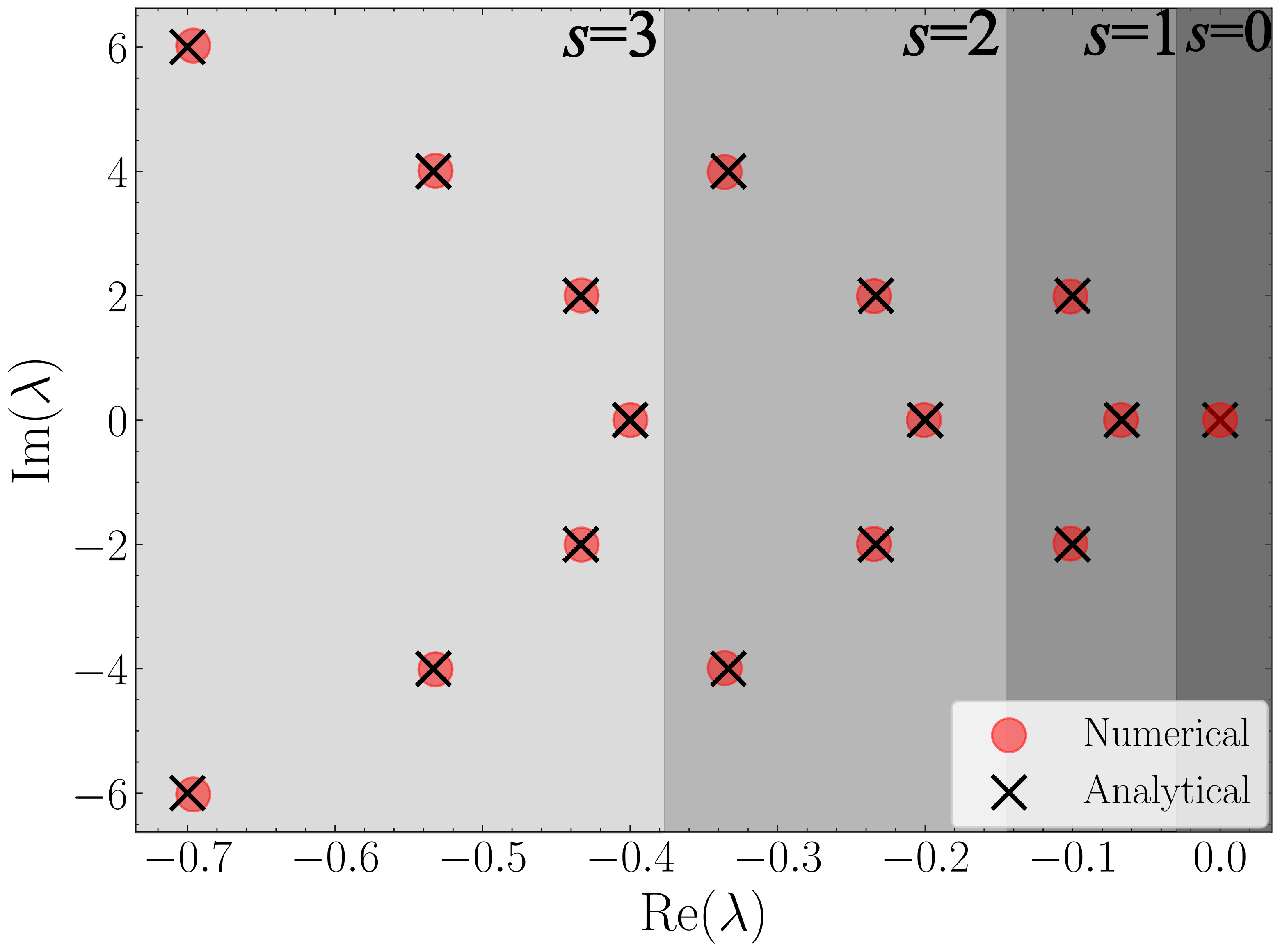}
    \caption{The Liouvillian spectrum for the BTC model given in Eq.~(\ref{eq:nori_model}), showing a comparison between the numerically exact results (red circles) and the analytical results obtained via the superspin method (black crosses). Here $N=3$, $\Omega_x=1$ and $\Gamma=0.1$.}
    \label{fig:liouvillian_spectrum_btc}
\end{figure}

Written in superoperator form as in Eq.~(\ref{definition_L0}), the coherent dynamics may be reinterpreted as that of a non-interacting bipartite system. It describes a subsystem governed by a non-Hermitian Hamiltonian, $-iH_S$, and a dual subsystem by $iH_S^T$. These will be referred to as the unprimed and primed systems, respectively. In the case of the BTC model of Eq.~(\ref{eq:nori_model}), the coherent term satisfies the eigenvalue equation
\begin{equation}
    \mathcal{L}_0 \ket{m_x, m'_x}\rangle = 2i\Omega_x(m_x-m'_x) \ket{m_x, m'_x}\rangle,
    \label{eigenval_eqn_subsystems}
\end{equation}
where we have chosen to use the $J_x$ eigenbasis such that the full eigenstates are $\ket{j, m_x; j, m'_x}$. To keep the notation compact, the $j$ dependence has been dropped. The coherent term describes a highly degenerate system since the eigenvalues only depend on the difference \mbox{$m_x-m'_x$}. Thus, it is convenient to introduce a new operator $S_x=J_x \otimes\mathbb{I}- \mathbb{I}\otimes J_x^T$ whose eigenvalues are $(2/N)s_x$ with~$s_x=m_x-m'_x$. This operator is the $x$-projection of another operator which we term the \textit{superspin}, $\mathbf{S}=\mathbf{J}\otimes\mathbb{I}-\mathbb{I}\otimes \mathbf{J}^T$. It represents the relative spin between the two subsystems. Then, the superspin projections are given by 
\begin{equation}
    S_{\alpha}= J_{\alpha}\otimes \mathbb{I}- \mathbb{I}\otimes J_{\alpha}^T
\end{equation}
for $\alpha=x,y,z$. Employing the superspin operator is analogous to using the total angular momentum operator when dealing with angular momentum addition in quantum mechanics. However, here the tensor product structure is a consequence of the superoperator formalism. Both the unprimed and primed subsystems represent the same underlying spin system, but we make use of the mathematical structure of $\mathcal{L}_0$ to reinterpret the dynamics as that of two spins whose angular momenta are subtracted. We define the operator $S^2=S_x^2 + S_y^2+S_z^2$ such that its eigenvalues are $(2/N)^2 s(s+1)$ in analogy with our collective spin operators. Spin addition implies the allowed values of $s$ are $0, ..., N$ in integer steps, where we made use of the fact that $j=N/2$.

We can trivially show that $S_x$ commutes with $J_x\otimes \mathbb{I}, \mathbb{I} \otimes J_x^T, J^2\otimes \mathbb{I}$ and $\mathbb{I}\otimes J^2$. These operators form a mutually commuting set of operators and therefore, we may label this basis as $\ket{m_x, m'_x,s_x}\rangle$, dropping the $j$ dependence. Alternatively, we can prove that $[S^2, S_x]=0$ and thus, label our basis as $\ket{s, s_x}\rangle$. For a proof of this result, see \hyperref[appendix:coupled_superspin_basis]{Appendix} \ref{appendix:coupled_superspin_basis}.
As a result, one has two options: either work in the uncoupled basis, focusing on the subsystems, or in the coupled (superspin) basis, treating the unprimed and primed systems as a single system. In either case, Eq.~(\ref{eigenval_eqn_subsystems}) becomes
\begin{equation}
    \mathcal{L}_0 \ket{\{\alpha\}, s_x}\rangle = 2i\Omega_x s_x \ket{\{\alpha\}, s_x}\rangle,
\end{equation}
where $\{\alpha\}$ represents the set of good quantum numbers. For the BTC model, we find that the dissipative dynamics simplifies if one works in the superspin basis.

To compute the first-order corrections to the eigenvalues, one needs to diagonalise the perturbation, $\mathcal{L}_D$, in the degenerate subspaces, which are labelled by $s_x$. Therefore, we need to compute the matrix elements $\mathcal{L}_D^{\alpha, \beta} = \langle\bra{\{\alpha\}, s_x}\mathcal{L}_D \ket{\{\beta\}, s_x}\rangle$. For the coupled basis, this results in $\mathcal{L}_D^{\tilde{s}, s} = \langle\bra{\tilde{s}, s_x }\mathcal{L}_D \ket{s, s_x}\rangle$. The perturbation for the BTC model in Eq.~(\ref{eq:nori_model}) is given by
\begin{equation}
\begin{split}
    \mathcal{L}_D
     = N \Gamma \Bigl(J_- \otimes J_-^*
    - \frac{1}{2}(J_+J_- \otimes \mathbb{I} + \mathbb{I}\otimes (J_+J_-)^T)\Bigr). \\
\end{split}
\end{equation}
Since the perturbation needs to be evaluated in the degenerate subspaces, terms that change the value of $s_x$ do not contribute. The terms that do contribute simplify to
$-(N\Gamma /4) (S_x^2+ S^2)$ (see \hyperref[appendix:derivation_eff_liouvillian_btc_model]{Appendix} \ref{appendix:derivation_eff_liouvillian_btc_model}), implying that
\begin{equation}
    \mathcal{L}_D^{\tilde{s}, s} = -\frac{\Gamma}{N}(s_x^2 + s(s+1))\delta_{\tilde{s}s}
    \label{btc_eig_diagonal}
\end{equation}
and so the perturbation is already diagonal in each degenerate subspace when expressed in the superspin basis. First-order shifts can be read-off from this expression as $\lambda_{s, s_x}^{(1)}=-(\Gamma/N )(s_x^2+s(s+1))$. Thus, to first-order, the eigenvalues are 
\begin{equation}
    \lambda_{s,s_x}= 2i\Omega_x s_x - \frac{\Gamma}{N} (s_x^2+s(s+1)).
    \label{btc_eigenvalues}
\end{equation}

Note that these results are consistent with the perturbative analysis of Ref.~\cite{btcs_pt_nakanishi}, where the first-order correction was formulated as a tridiagonal matrix in the uncoupled basis and used to establish the qualitative spectral properties of the Liouvillian, such as the emergence of purely imaginary eigenvalues and the closing of the Liouvillian gap in the thermodynamic limit. In that approach, however, the perturbative Liouvillian must, in general, be diagonalised numerically within degenerate subspaces, since explicit closed-form expressions for the resulting tridiagonal matrices are not available except in special cases.

\begin{figure}[t]
    \centering
    \includegraphics[width=\linewidth]{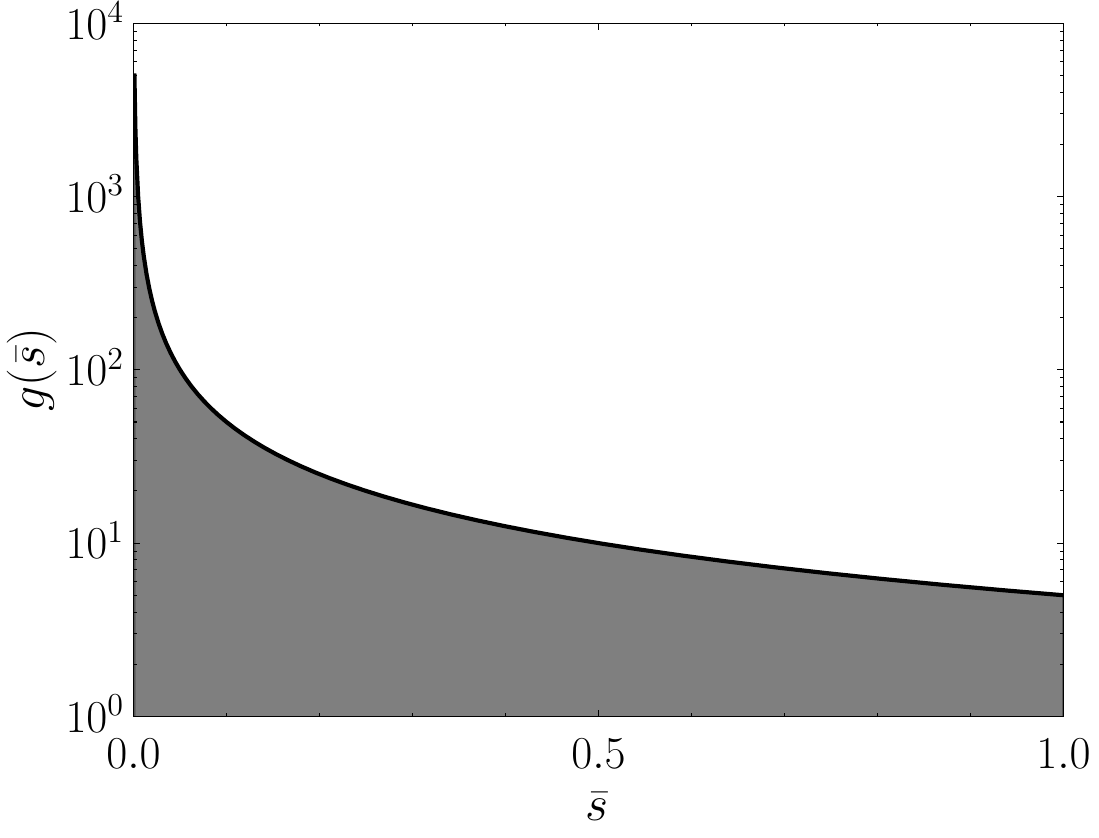}
    \caption{An illustration of the thermodynamic limit of $g(\bar{s})$ given in Eq.~(\ref{dos_eqn}) for $\Gamma=0.1$. We observe that the sectors become densely populated near $\bar{s}=0$. }
    \label{fig:sector_density}
\end{figure}

In contrast, by working directly in the superspin representation of Liouville space, we identify the symmetry-resolved basis in which the perturbative Liouvillian is diagonal from the outset. This allows us to derive explicit closed-form expressions for the Liouvillian eigenvalues, given in Eq.~(\ref{btc_eigenvalues}), thereby resolving the detailed spectral structure underlying the qualitative features identified previously. These expressions quantitatively account for the Liouvillian spectrum shown in Fig.~\ref{fig:liouvillian_spectrum_btc}.
The spectrum has distinct sectors, which arise due to the many allowed values of $s$ ($s \in \{0, 1, ..., N\}$). Within each sector, we see a quadratic dependence, which is due to the real parts being dependent on $s_x^2$, and this is proportional to $\mathrm{Im}(\lambda)^2$. In addition, for each sector, the allowed values of $s_x$ are $-s\leq s_x \leq s$. In the thermodynamic limit $N\to \infty$, sectors become denser and eigenvalues get arbitrarily close to the imaginary axis (see below for a quantitative discussion). This in turn implies that the system undergoes persistent oscillations of period determined by the eigenvalues of Eq.~(\ref{btc_eigenvalues}) with the smallest real part and nontrivial imaginary part, {\it i.e.} $s=1$ and $s_x = \pm 1$.

To be more quantitative, as Fig.~\ref{fig:liouvillian_spectrum_btc} shows, the distance between sectors decreases as we go from high to low $s$. We can quantify this by defining the distance between neighbouring sectors, $d(s)$, as 
\begin{equation}
    d(s)=\abs{ \lambda_{s,0} - \lambda_{s-1, 0}}.
\end{equation}
Here, $\lambda_{s,0}$ corresponds to the distance along the real axis between the origin and the eigenvalue with $s_x=0$ in the sector labelled by $s$. The sectors become densely populated in the thermodynamic limit. It is therefore useful to define the density of sectors as the number of sectors per unit distance. Using the expression given in Eq.~(\ref{btc_eigenvalues}), we obtain that (see Fig.~\ref{fig:sector_density})
\begin{equation}
    g(\bar{s})=\frac{1}{d(\bar{s})}=\frac{1}{2\Gamma\bar{s}},
    \label{dos_eqn}
\end{equation}
where $g(\bar{s})$ is the density of sectors and $\bar{s}=s/N$ such that $\bar{s}\in[0,1]$. In the thermodynamic limit, the low-$\bar{s}$ region of the Liouvillian spectrum becomes densely populated. The density of sectors diverges as $\bar{s} \to 0^{+}$. As a result, we observe a macroscopic accumulation of sectors with vanishing real components.

\begin{figure}[t]
    \centering
    \includegraphics[width=1\linewidth]{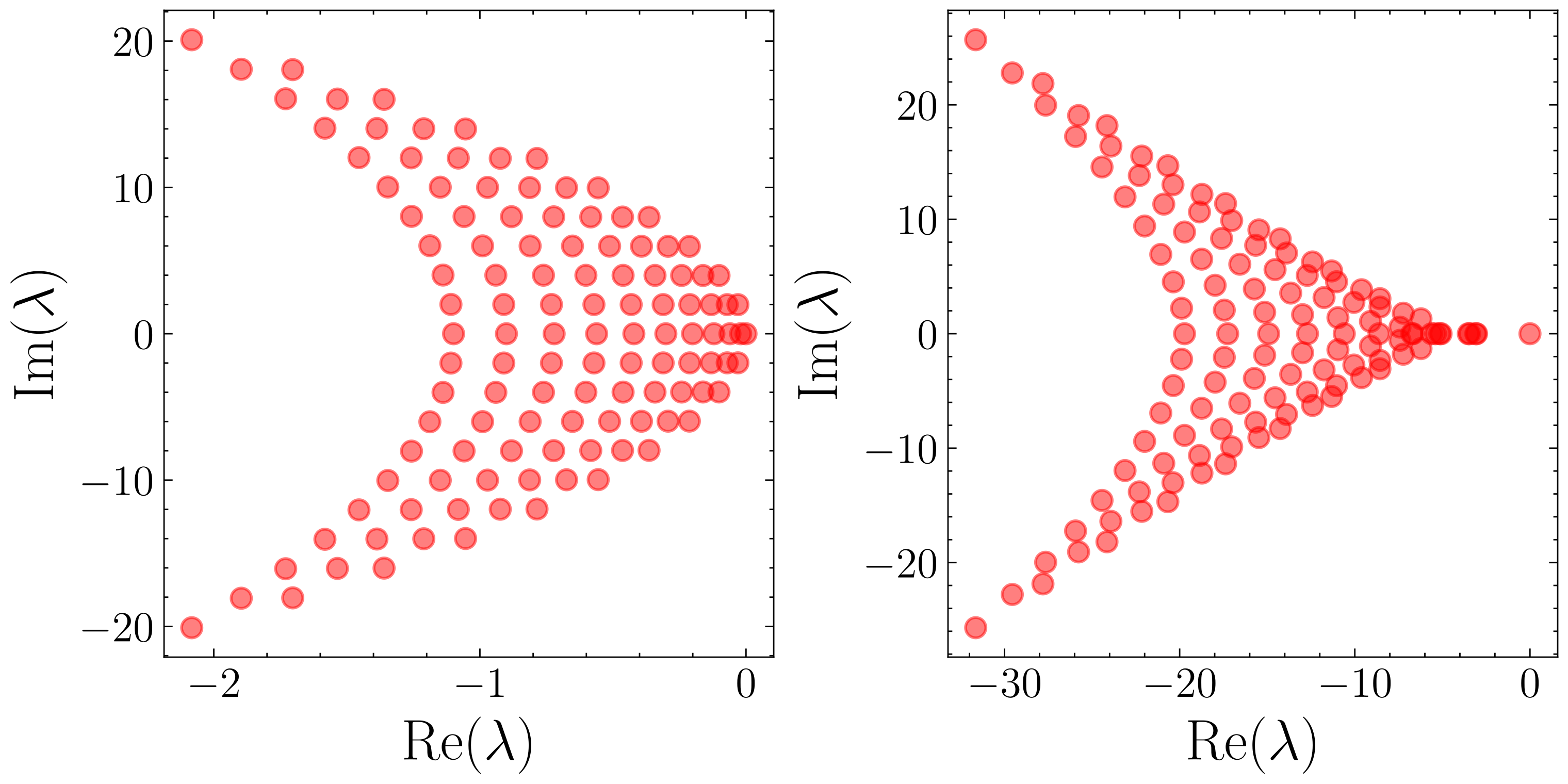}
    \caption{Liouvillian spectra for the BTC model of Eq.~(\ref{eq:nori_model}) in the case of $N=10$ and $\Omega_x=1$. On the left we show the spectrum for $\Gamma=0.1$ and on the right for $\Gamma=2$. The numerically computed eigenvalues are represented via red circles.}
    \label{fig:btc_model_phase_transition}
\end{figure}

Consequently, a large subset of sectors hosts eigenvalues with vanishing real parts, resulting in a closure of the Liouvillian gap. This leads to a highly degenerate nonequilibrium steady-state subspace, signalling the emergence of a BTC phase. Moreover, within these sectors, eigenvalues with non-zero $s_x$ retain a finite imaginary component, giving rise to long-lived coherences with an infinite timescale and, consequently, persistent oscillations.

This model undergoes a BTC to stationary steady-state phase transition as $N\to \infty$ at $\Gamma = \Omega_x$ \cite{BTC1,franco_nori_btc}, which can be understood by analysing the eigenvalue spectrum. Considering the finite-$N$ case, as illustrated in Fig. \ref{fig:btc_model_phase_transition}, we observe that as $\Gamma$ increases, the eigenvalues progressively collapse onto the real axis, resulting in a gapped spectrum. Within a given sector, complex conjugate eigenvalue pairs associated with the same $|s_x|$ merge at exceptional points. Notably, this transition occurs gradually, first affecting the eigenvalue pairs with the smallest negative real parts. Since these eigenvalues govern the persistent oscillations in the thermodynamic limit, their coalescence signals the destruction of the BTC phase, as oscillatory behaviour ceases when the eigenvalues become purely real. Consequently, this model is highly sensitive to increasing dissipation strength, and the BTC phase is unstable against increasing dissipation.
The perturbative approach remains valid as long as $\Gamma$ remains small. To describe the merging of eigenvalues, we need higher-order corrections, which must have an imaginary part to account for the vanishing imaginary components. This is easily illustrated by the $N=1$ case, where the problem can be solved exactly. In this case, we have two sectors, $s=0$ and $s=1$. The $s=1$ sector contains three eigenvalues, which are $\lambda=-2\Gamma, -3\Gamma \pm 2i \Omega_x \sqrt{1-(\Gamma/2\Omega_x)^2}$. As these expressions indicate, shifts in the imaginary components are at least second-order corrections.

\section{Non-BTC Collective Spin Models}
\label{sec:other_models}
\begin{figure}[t]
    \centering
    \includegraphics[width=1\linewidth]{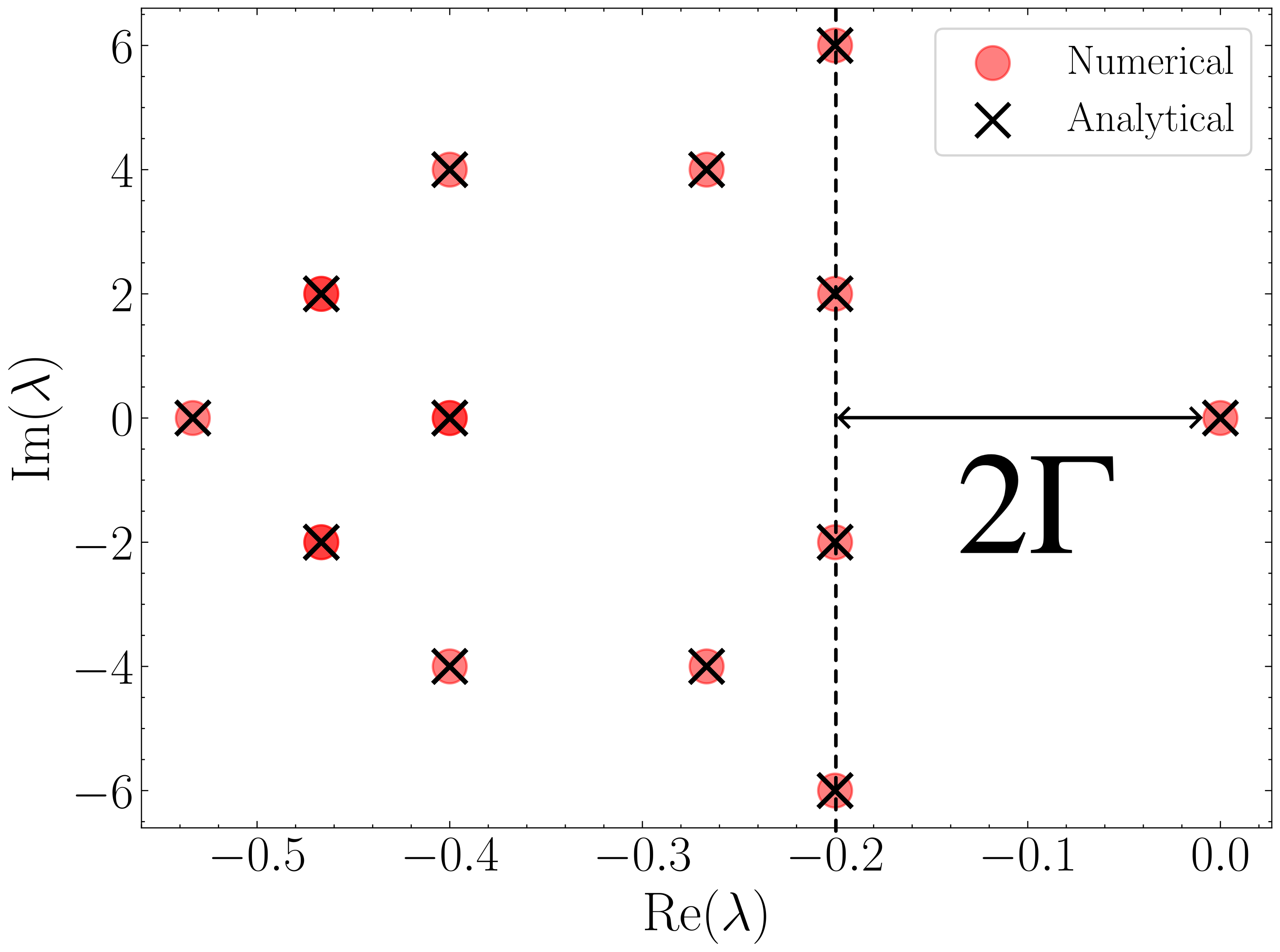}
    \caption{The Liouvillian spectrum for model A, showing a comparison between the numerically exact results (red circles) and the analytical results obtained via the superspin method (black crosses). Here $N=3$, $\Omega_x=1$ and $\Gamma=0.1$.}
    \label{fig:model_a_eigenvalue_spectrum}
\end{figure}

\subsection{Model A}
Next, we demonstrate how the perturbative method allows one to study other spin models and analyse their eigenvalue spectra. Let us first consider a model where $H_S=-N\Omega_z J_z$ and we have a single Lindblad jump operator, $J_+$ (in the $z$-basis), with a corresponding rate $N\Gamma$. In this case, we use the uncoupled basis as the perturbation cannot be expressed purely in terms of the superspin. The perturbation may be written as
\begin{equation}
\begin{split}
    \mathcal{L}_D 
    & = N \Gamma J_+ \otimes J_+^* - \frac{N\Gamma}{2}\Bigl[(J^2-J_z^2-\frac{2}{N}J_z) \otimes \mathbb{I}\\
    &\hspace{90pt}+ \mathbb{I}\otimes(J^2 -J_z^2 -\frac{2}{N}J_z)\Bigr],\\
\end{split}
\end{equation}
which becomes an upper triangular matrix in the degenerate subspaces. The eigenvalues of upper triangular matrices are given by the diagonal elements, meaning $J_+\otimes J_+^*$ does not affect the eigenvalues of $\mathcal{L}_D$. Thus, the eigenvalues are
\begin{equation}
\begin{split}
    \lambda_{m_z, m'_z}
    & = 2i \Omega_z (m_z-m'_z)\\
    & - \frac{2\Gamma}{N}\Bigl(N \Bigl(\frac{N}{2}+1\Bigr) -m_z(m_z+1) -m'_z(m'_z+1)\Bigr),\\
\end{split}
\end{equation}
with the eigenvalue spectrum shown in Fig. \ref{fig:model_a_eigenvalue_spectrum}.
For this model, the minimum Liouvillian gap can be calculated via $\abs{\mathrm{Re}(\lambda_{m_z, m'_z})}$ with $(m_z, m'_z)=(N/2, N/2-1)$ and is equal to the constant $2\Gamma$. As a result, the gap does not vanish in the thermodynamic limit, and therefore, we do not observe persistent oscillations and a BTC phase.

\subsection{Model B}
\begin{figure}[t]
    \centering
    \includegraphics[width=1\linewidth]{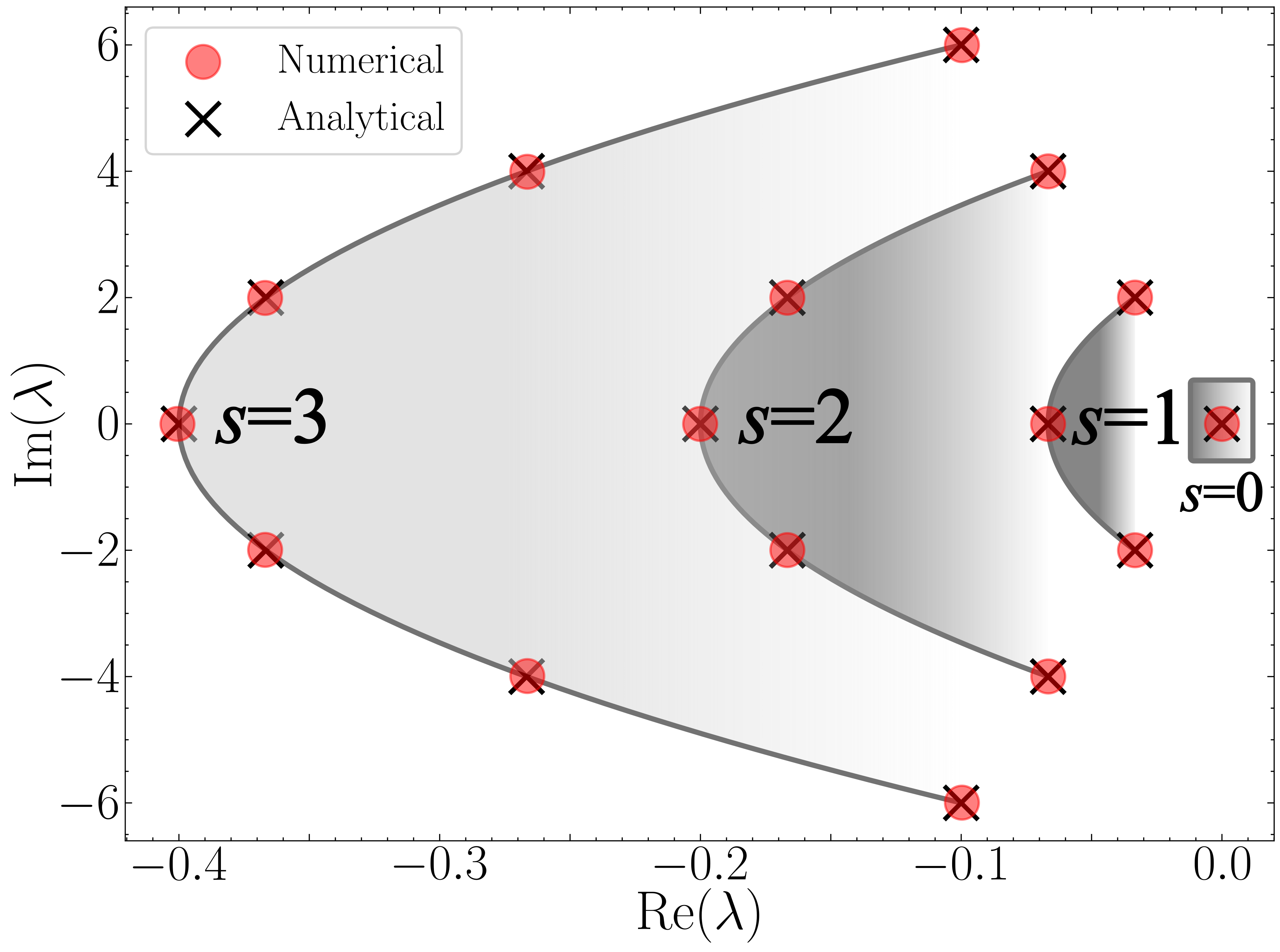}
    \caption{The Liouvillian spectrum for model B, showing a comparison between the numerically exact results (red circles) and the analytical results obtained via the superspin method (black crosses). Here $N=3$, $\Omega_x=1$ and $\Gamma=0.1$.}
    \label{fig:model_b_eigenvalue_spectrum}
\end{figure}

Let us consider another example, where we do indeed observe a vanishing gap in the real parts. Consider the case where $H_S=-N\Omega_x J_x$ but now the Lindblad jump operator is $J_z$, with rate $N\Gamma$. The perturbation is given by 
\begin{equation}
    \mathcal{L}_D = N\Gamma \Bigl(J_z \otimes J_z^* - \frac{1}{2}(J_z^2 \otimes \mathbb{I} + \mathbb{I}\otimes J_z^2)\Bigr),
\end{equation}
where $J_z=(1/2i)(J_+^{(x)}- J_-^{(x)})$. Following the method outlined in \hyperref[appendix:derivation_eff_liouvillian_btc_model]{Appendix} \ref{appendix:derivation_eff_liouvillian_btc_model}, in the superspin basis we obtain
\begin{equation}
    \mathcal{L}_D^{\tilde{s},s}=\frac{\Gamma}{N}(s_x^2-s(s+1))\delta_{\tilde{s}s}.
\end{equation}
Thus, to first-order in $\Gamma$, the eigenvalues are
\begin{equation}
    \lambda_{s,s_x}=2i\Omega_x s_x +\frac{\Gamma}{N}(s_x^2-s(s+1)),
\end{equation}
as shown in Fig. \ref{fig:model_b_eigenvalue_spectrum}.
Once again, due to the many allowed values of $s$ and the $1/N$ scaling, we observe a vanishing Liouvillian gap and vanishing real parts when the number of sectors becomes macroscopic. The spectrum is almost identical to that of the BTC model of Eq.~(\ref{eq:nori_model}), except for the sign of the quadratic term ($s_x^2$).

\begin{figure}[t]
    \centering
    \includegraphics[width=1\linewidth]{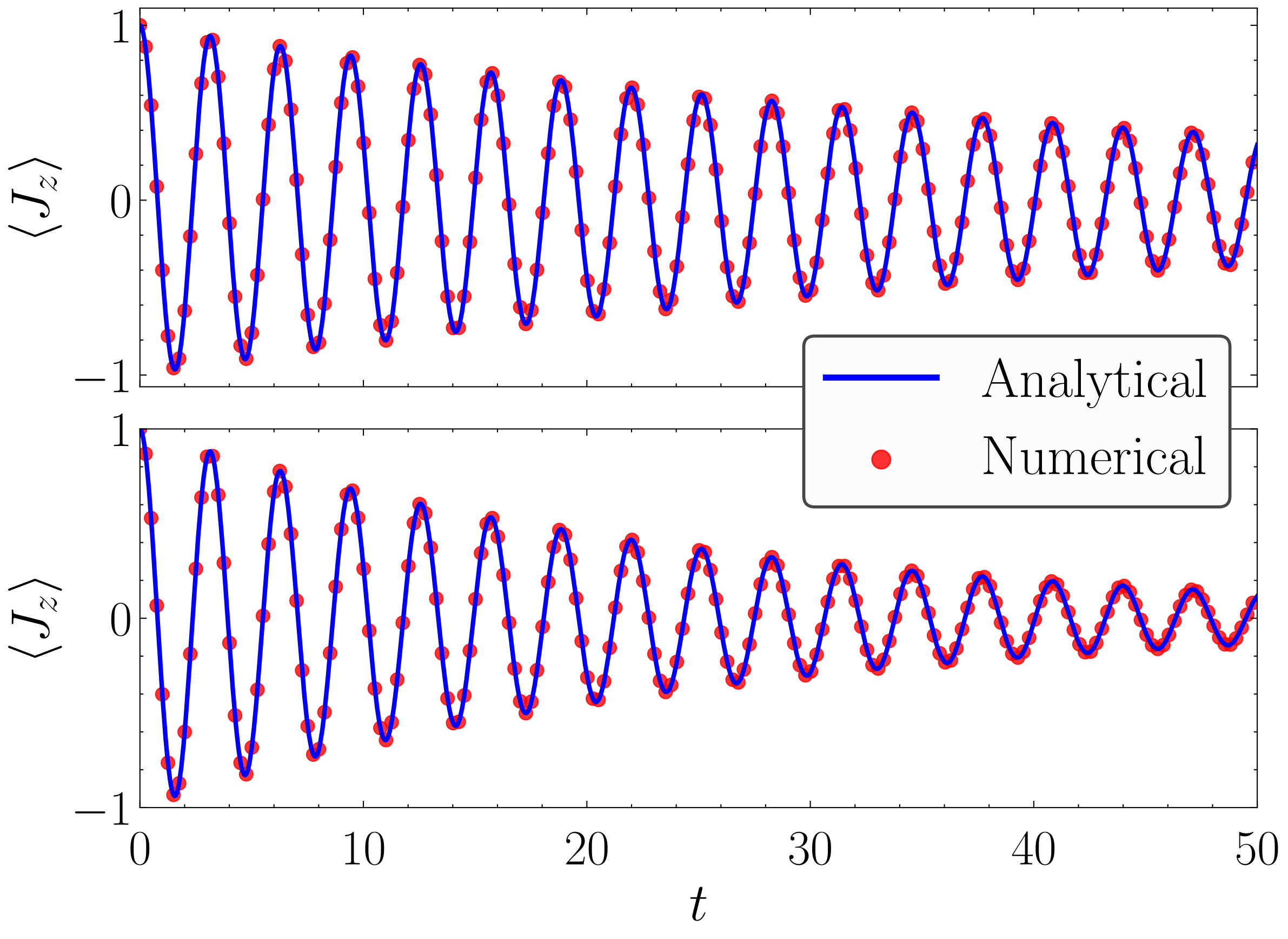}
    \caption{Decaying oscillations in $\langle J_z(t)\rangle$ as a function of time, $t$, for models B and C are shown on the top and bottom, respectively. For both cases, $N=100$, $\Omega_x=1$, $\Gamma=2$ and the system is initialised in a pure state in the $J_z$ basis such that \mbox{$\langle J_z(0)\rangle=1$}. Numerical solutions are shown via red circles and analytical solutions via blue lines.}
    \label{fig:JxJz_JxJx_oscillations}
\end{figure}

\begin{figure}[t]
    \centering
    \includegraphics[width=1\linewidth]{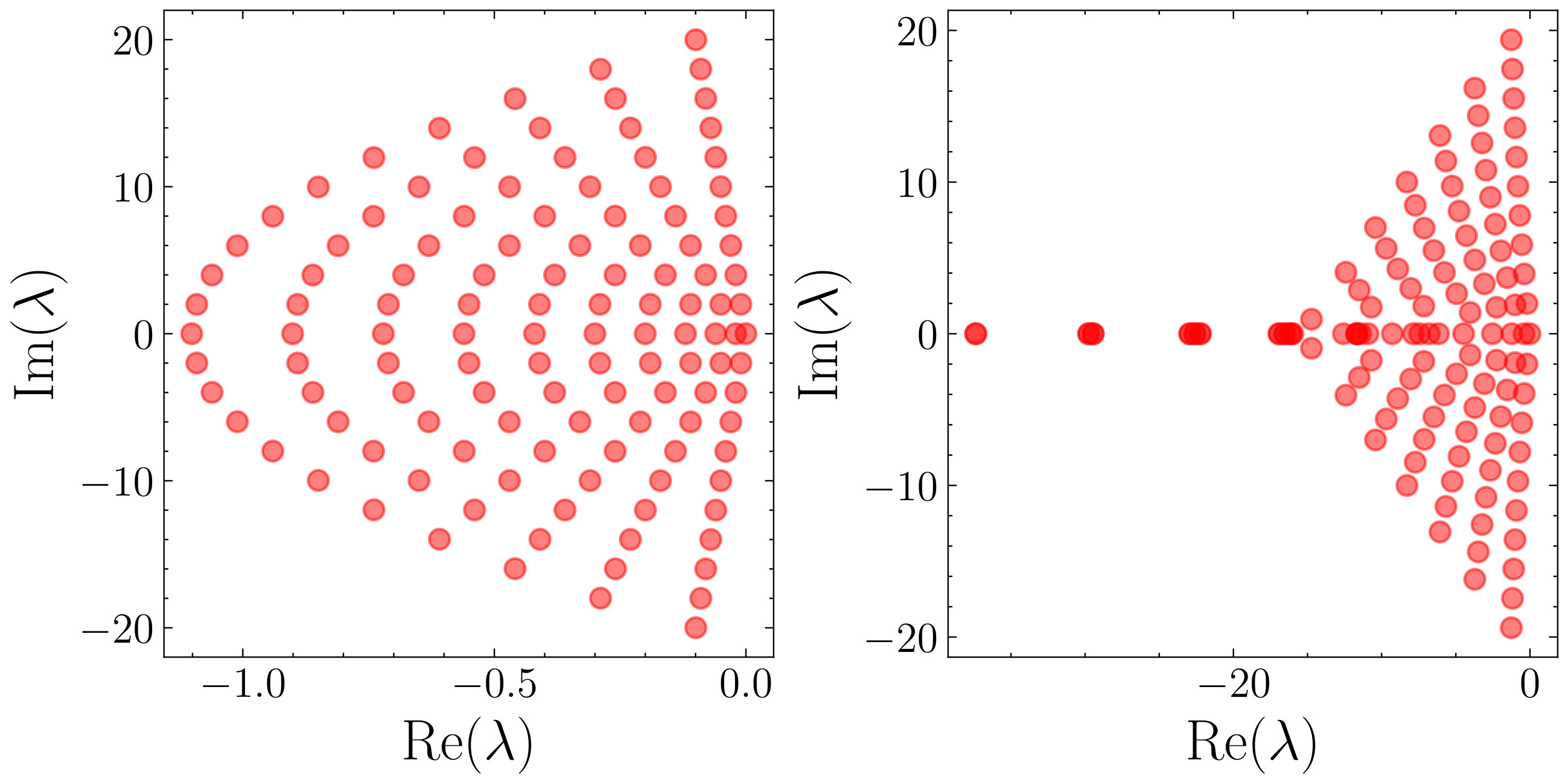}
    \caption{Liouvillian spectra for model B in the case of $N=10$ and $\Omega_x=1$. On the left we show the spectrum for $\Gamma=0.1$ and on the right for $\Gamma=2$. The numerically computed eigenvalues are represented via red circles.}
    \label{fig:JxJz_phase_transition}
\end{figure}

Model B can also lead to decaying oscillations in $\langle J_z(t)\rangle$ (in the finite-$N$ case) as shown on the top of Fig.~\ref{fig:JxJz_JxJx_oscillations}. In contrast to the BTC model of Eq.~(\ref{eq:nori_model}), the transition from persistent oscillations to decay towards a uniform steady state here begins with the eigenvalues that have the most negative real parts (see Fig.~\ref{fig:JxJz_phase_transition}). This transition unfolds gradually, following the same mechanism as before: complex conjugate eigenvalue pairs within the same sectors merge, creating exceptional points. However, since this transition starts from the opposite side, this model is more robust against increasing dissipation. In the thermodynamic limit, the transition must first affect a macroscopic number of sectors starting from \( s = N \) before reaching those with the least negative real parts. As a result, the sectors responsible for the persistent oscillations cannot undergo this transition, making this behaviour stable against increasing dissipation.

We can assess whether these persistent oscillations correspond to genuine BTC behaviour by analysing the dynamics of collective observables using the Ehrenfest equations of motion \cite{breuer_open_qm}. These equations allow us to solve for the expectation values $\langle J_{\alpha}(t) \rangle$ exactly, without relying on a perturbative expansion in the dissipation strength $\Gamma$.

Following the method outlined in Appendix~\ref{appendix:evolution_eqn_spin_operators}, we find that the dynamics of $\langle J_z \rangle$ is governed by
\begin{equation}
    \frac{d^2}{dt^2} \langle J_z \rangle + 2\kappa \frac{d}{dt} \langle J_z \rangle + \omega^2 \langle J_z \rangle = 0,
\end{equation}
where $\kappa = \Gamma/N$ and $\omega = 2\Omega_x$. This yields the solution
\begin{equation}
    \langle J_z(t)\rangle =
    \Bigl(
        c_1 \cos(f(\omega,\kappa)t)
        + c_2 \sin(f(\omega,\kappa)t)
    \Bigr)\,
    \mathrm{e}^{-\kappa t},
\end{equation}
with $f(\omega,\kappa) = \omega \sqrt{1 - (\kappa/\omega)^2}$ and constants $c_1,c_2$ determined by the initial conditions.

This solution describes damped harmonic motion characterised by a \emph{single} oscillation frequency $f(\omega,\kappa)$. Crucially, this behaviour is fundamentally different from that of a genuine BTC, where oscillatory dynamics arises from a ladder of frequencies, producing multiple sharp, evenly spaced spectral lines in the Fourier spectrum of $\langle J_z \rangle$ \cite{BTC1}. No such multifrequency structure is present here.

Moreover, as shown in Appendix~\ref{appendix:evolution_eqn_spin_operators}, $\langle J_x \rangle$ decouples from the remaining spin components, implying that the observed oscillations depend sensitively on the choice of initial state rather than emerging as a collective dynamical order. The resulting dynamics closely resembles single-spin Rabi oscillations \cite{Booker_2020}, rather than time-crystalline behaviour.

We therefore conclude that, although this model exhibits persistent oscillations in the thermodynamic limit, it does \emph{not} realise a genuine BTC phase.

\subsection{Model C}

\begin{figure}[t]
    \centering
    \includegraphics[width=\linewidth]{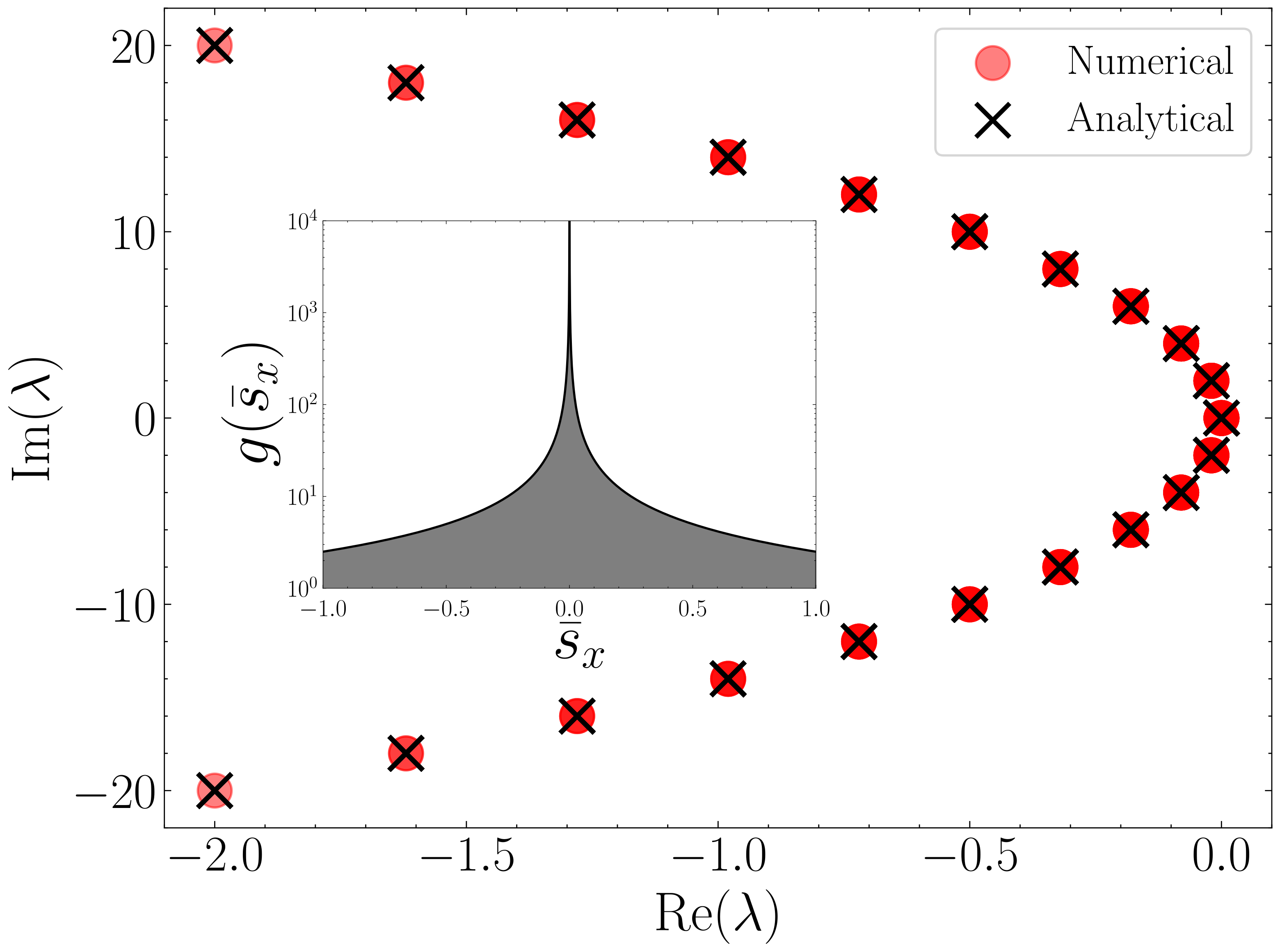}
    \caption{The Liouvillian spectrum for model C in the case of $N=10$, $\Omega_x=1$ and $\Gamma=0.1$. Numerically exact eigenvalues are shown via red circles and analytical results via black crosses. An inset shows $g(\bar{s}_x)$, illustrating how the eigenmodes are densely populated near $\bar{s}_x=0$ in the thermodynamic limit.}
    \label{fig:JxJx_with_inset}
\end{figure}

A pure-dephasing model can be described by $H_S=-N\Omega_x J_x$ and a single Lindblad jump operator, $J_x$, with rate $N\Gamma$. Here, the incoherent term is given by
\begin{equation}
    \mathcal{L}_D = N\Gamma \Bigl(J_x \otimes J_x - \frac{1}{2}(J_x^2 \otimes \mathbb{I}+ \mathbb{I}\otimes J_x^2)\Bigr).
\end{equation}
This model can be solved exactly, with its eigenvalues given by
\begin{equation}
    \lambda_{s_x}= 2i \Omega_x s_x - \frac{2\Gamma}{N}s_x^2,
\end{equation}
which does not depend on $s$ and hence, does not show splitting due to different sectors (see Fig. \ref{fig:JxJx_with_inset}). The values of $s_x$ are in the range $-N\leq s_x \leq N$ in integer steps. In this case, we can define the distance along the real axis between adjacent eigenmodes, $d(s_x)$, as
\begin{equation}
    d(s_x)=\abs{\mathrm{Re}(\lambda_{s_x})- \mathrm{Re}(\lambda_{s_x-1})},
\end{equation}
which in the thermodynamic limit gives a density of eigenmodes, $g(\bar{s}_x)$, that satisfies
\begin{equation}
    g(\bar{s}_x) = \frac{1}{d(\bar{s}_x)}=\frac{1}{4\Gamma \abs{\bar{s}_x}}.
\end{equation}
Here, $\bar{s}_x=s_x/N$ such that $\bar{s}_x \in [-1,1]$. Qualitatively, the thermodynamic behaviour of this model is similar to that of the BTC model of Eq.~(\ref{eq:nori_model}), as shown in Fig.~\ref{fig:JxJx_with_inset}. We observe a sharp peak in $g(\bar{s}_x)$ as $\bar{s}_x\to 0^{\pm}$, indicating the presence of a macroscopic number of modes with vanishing real parts. Some of these modes have finite imaginary parts; therefore, this model also supports persistent oscillations.

Model~C exhibits the same qualitative dynamical behaviour as model~B when analysed at the level of collective observables. By considering the Ehrenfest equations of motion for $\langle J_{\alpha}(t) \rangle$, we find that the dynamics of $\langle J_z(t) \rangle$ is given by
\begin{equation}
    \langle J_z(t)\rangle =
    \Bigl(
        c_1 \cos(\omega t)
        + c_2 \sin(\omega t)
    \Bigr)\,
    \mathrm{e}^{-2\kappa t},
\end{equation}
where $c_1$ and $c_2$ are constants fixed by the initial conditions. This behaviour is illustrated in the bottom panel of Fig.~\ref{fig:JxJz_JxJx_oscillations}.

As in model~B, the equations of motion reduce to those of a damped harmonic oscillator characterised by a \emph{single} oscillation frequency. This is in sharp contrast to genuine BTC dynamics, which originates from a ladder of frequencies and manifests as multifrequency oscillations in time-dependent observables. No such multifrequency structure is present in model~C.

We therefore conclude that, despite exhibiting persistent oscillations in the thermodynamic limit, model~C does \emph{not} realise a genuine BTC phase. The oscillatory dynamics instead reflects single-frequency collective motion.

\section{Discussion and Conclusions}
\label{sec:conclusions}

In this paper, we have introduced an analytical approach to investigate the microscopic Liouvillian eigenvalue spectrum of the canonical BTC model. By treating the dissipation perturbatively, we reinterpret the dynamics of the collective spin system in terms of two fictitious subsystems, which can be combined into an effective ``superspin’’ degree of freedom. Working in this superspin representation allows us to derive explicit closed-form expressions for the eigenvalues of the effective Liouvillian governing the long-time dynamics in the extreme time-crystal phase.

For the paradigmatic BTC model [Eq.(\ref{eq:nori_model})], the spectrum organises into superspin multiplets with quadratic profiles. Taking the thermodynamic limit of these eigenvalues reveals the formation of a dense set of modes near the origin of the complex plane: the separation between adjacent superspin sectors along the real axis vanishes as the system size increases, while the imaginary parts remain finite. As a result, a macroscopic number of Liouvillian eigenvalues acquire vanishingly small real parts and finite imaginary parts in the thermodynamic limit. This provides a microscopic spectral description of the BTC phase and clarifies the mechanism underlying spontaneous breaking of continuous time-translation symmetry in the model defined by Eq.(\ref{eq:nori_model}).

{We subsequently applied the same analytical framework to other dissipative spin models, including model B, which was previously analysed in Ref.~\cite{btcs_pt_nakanishi}. While that work argued that balanced gain and loss at first order is sufficient to generate BTC behaviour, our results show that this criterion alone is not sufficient. By analysing the exact equations of motion for collective spin observables, we demonstrate that both models B and C reduce to damped harmonic motion characterised by a single oscillation frequency. This single-frequency collective behaviour is fundamentally distinct from BTC oscillations, which arise from a multifrequency structure.

Our analysis therefore establishes that Liouvillian spectral properties alone do not constitute definitive evidence of BTC dynamics: persistent oscillations and gapless spectra must be accompanied by the appropriate microscopic dynamical structure. Although models B and C do exhibit persistent oscillations in the thermodynamic limit, these oscillations reflect single-frequency collective dynamics rather than genuine BTC behaviour. Clarifying the broader classification of oscillatory but non-BTC dissipative dynamics, and their relation to genuine time-crystal order, remains an interesting direction for future work.

\begin{acknowledgments}
D.N. acknowledges E. Asquith and S. Meredith for valuable discussions. This work was supported by the Dean's Doctoral Scholarship from The University of Manchester.
A.P. and D.N. acknowledge support from the European Commission under the EU Horizon 2020 MSCA-RISE-2019 programme (project 873028 HYDROTRONICS). A.P. also acknowledges support from the Leverhulme Trust under the grant agreement RPG-2023-253.
\end{acknowledgments}

\section*{Code Availability}
The Python code used to generate the results in this paper is available at \cite{nemeth2025btc-code}.

\appendix
\section{Properties of Collective Spin Operators}
\label{appendix:collective_spin_operators}
We consider a collective spin model where the total spin operator is constructed from the individual spins such that
\begin{equation}
\begin{split}
    \mathbf{J} 
    & = \frac{2}{N}\sum_{i=1}^N \boldsymbol{\sigma}^{(i)} / 2\\
    &= \frac{2}{N}(\boldsymbol{\sigma}^{(1)}/2\otimes \mathbb{I}^{(2)}...\otimes \mathbb{I}^{(N)} + ... \\
    &\, \, \, \, \,+ \mathbb{I}^{(1)}\otimes...\otimes \mathbb{I}^{(N-1)} \otimes \boldsymbol{\sigma}^{(N)}/2),\\
\end{split}
\end{equation}
where $\mathbb{I}^{(i)}$ is the identity operator acting on the $i^{\mathrm{th}}$ site. This normalisation allows one to interpret $\{J_{\alpha}\}$  as a set of magnetisation operators.

In some cases, it is convenient to work in the eigenbasis of $J_x$. Then, eigenstates of $J_x$ and $J^2$ satisfy
\begin{equation}
    J_x \ket{j, m_x} = \frac{2}{N}m_x \ket{j, m_x},
\end{equation}
where $m_x =-N/2, ..., N/2$ in integer steps. We can cyclically permute the spin operators such that $J_z \to J_x$, $J_y\to J_z$ and $J_x\to J_y$, allowing us to define raising and lowering operators in the $J_x$ basis as $J_\pm^{(x)}= J_y \pm i J_z$. These raising and lowering operators satisfy $[J_x, J_\pm^{(x)}]=\pm (2/N)J_\pm^{(x)}$. An important consequence of these definitions is the relation
\begin{equation}
\begin{split}
    J_{\pm}
    & =J_x\pm iJ_y \\
    & = J_x \pm \frac{i}{2}(J_+^{(x)}+J_-^{(x)})
    \label{J_pm to J_pm^(x)}
\end{split}
\end{equation}
and raising and lowering operators in the $x$-basis satisfy
\begin{equation}
    J_{\mp}^{(x)}J_{\pm}^{(x)} = J^2 - J_x^2 \mp\frac{2}{N}J_x.
    \label{J_pm^x expression with Jx}
\end{equation}

\section{Coupled Superspin Basis}
\label{appendix:coupled_superspin_basis}
To determine $[S^2, J_x \otimes \mathbb{I}], [S^2, \mathbb{I}\otimes J_x]$ we first need to find an expression for $S^2$. Taking the square of $\mathbf{S}$ we get
\begin{equation}
\begin{split}
    S^2
    & =(\mathbf{J}\otimes \mathbb{I} -\mathbb{I}\otimes \mathbf{J}^T)^2 \\
    & = J^2 \otimes \mathbb{I} + \mathbb{I} \otimes (J^T)^2 - 2 \mathbf{J} \otimes \mathbf{J}^T. \\
\end{split}
\end{equation}
Here, $(J^T)^2= (J_x^T)^2+(J_y^T)^2+(J_z^T)^2=J^2$, which can be shown using the fact that $J_{\alpha}$ is Hermitian and $J_\pm^{(x)}$ are real matrices in the basis where $J_x$ is diagonal. Rewriting the transposes again yields
\begin{equation}
\begin{split}
    \mathbf{J} \otimes \mathbf{J}^T 
    &= J_x\otimes J_x^T + J_y \otimes J_y^T + J_z \otimes J_z^T \\
     &= J_x\otimes J_x + J_y \otimes J_y - J_z \otimes J_z. \\
\end{split}
\end{equation}
The next step is to rewrite $J_y$ and $J_z$ in terms of $J_\pm^{(x)}$ so that
\begin{equation}
\begin{split}
 J_y \otimes J_y - J_z \otimes J_z
 & = \frac{1}{4}\Bigl((J_+^{(x)}+J_-^{(x)}) \otimes (J_+^{(x)}+J_-^{(x)})\\
 &\hspace{10pt}  + (J_+^{(x)}-J_-^{(x)}) \otimes (J_+^{(x)}-J_-^{(x)}) \Bigr)\\
   & = \frac{1}{2}\left(J_+^{(x)}\otimes J_+^{(x)} + J_-^{(x)} \otimes J_-^{(x)}\right).\\
\end{split}
\end{equation}
Finally, we obtain a simplified expression for $S^2$ such that
\begin{equation}
    S^2 = J^2 \otimes \mathbb{I} + \mathbb{I} \otimes J^2 -2 J_x \otimes J_x - \left(J_+^{(x)}\otimes J_+^{(x)} + J_-^{(x)} \otimes J_-^{(x)}\right). 
    \label{S2_defn}
\end{equation}
Now that we have this expression, we can determine which operators commute with $S^2$. Recall that in the $x$-basis $[J^2, J_x]=0$, which means that
\begin{equation}
\begin{split}
       [S^2, J_x \otimes I]
       & = - \left[\left(J_+^{(x)}\otimes J_+^{(x)} + J_-^{(x)} \otimes J_-^{(x)}\right), J_x\otimes I\right]\\
       & = - [J_+^{(x)}, J_x] \otimes J_+^{(x)} - [J_-^{(x)}, J_x] \otimes J_-^{(x)} \\
       & = \frac{2}{N}( J_+^{(x)} \otimes J_+^{(x)}) - \frac{2}{N} (J_-^{(x)} \otimes J_-^{(x)}) \\
       & \neq 0.
\end{split}
\end{equation}
Therefore, we cannot simultaneously diagonalise $S^2$ and $J_x \otimes \mathbb{I}$, and the same is true for $\mathbb{I} \otimes J_x$. However, using these results, we can show that $[S^2, S_x]=0$ since
\begin{equation}
\begin{split}
       [S^2, S_x]
       & = - [J_+^{(x)}, J_x] \otimes J_+^{(x)} - [J_-^{(x)}, J_x] \otimes J_-^{(x)} \\
       & \hspace{10pt} +J_+^{(x)}\otimes [J_+^{(x)}, J_x]  + J_-^{(x)}\otimes[J_-^{(x)}, J_x]\\
       & = \frac{2}{N}( J_+^{(x)} \otimes J_+^{(x)}) - \frac{2}{N} (J_-^{(x)} \otimes J_-^{(x)})\\
       & \hspace{10pt} - \frac{2}{N}( J_+^{(x)} \otimes J_+^{(x)}) + \frac{2}{N} (J_-^{(x)} \otimes J_-^{(x)})\\
       & = 0.
\end{split}
\end{equation}
We can understand this result more intuitively by focusing on Eq.~(\ref{S2_defn}). Raising (or lowering) the $x$-component of the spin states of both the unprimed and primed subsystems leaves $s_x$ invariant, therefore, $[J_+^{(x)} \otimes J_+^{(x)}, S_x]=[J_-^{(x)} \otimes J_-^{(x)}, S_x]=0$.

If $H_S$ includes $J_z$ instead of $J_x$, it is more instructive to work in the $J_z$ eigenbasis and hence express $J_x$ and $J_y$ in terms of raising and lowering operators of $J_z$. We can perform the same steps as above but using $J_\pm=J_x \pm i J_y$, leading to 
\begin{equation}
    S^2 = J^2 \otimes \mathbb{I} + \mathbb{I} \otimes J^2 -2 J_z \otimes J_z - \left(J_+\otimes J_+ + J_- \otimes J_-\right). 
\end{equation}
Therefore, the coupled basis is now given by the set $\{\ket{s, s_z}\rangle\}$.

\section{Derivation of the Effective Liouvillian for the BTC Model}
\label{appendix:derivation_eff_liouvillian_btc_model}
Our aim is to simplify the perturbation 
\begin{equation}
\begin{split}
    \mathcal{L}_D
    & = N \Gamma \Bigl(J_- \otimes J_-^*  - \frac{1}{2}(J_+J_- \otimes \mathbb{I} + \mathbb{I}\otimes (J_+J_-)^T)\Bigr). \\
\end{split}
\end{equation}
by only keeping terms that leave $s_x$ invariant. First, we rewrite all terms by invoking Eq.~(\ref{J_pm to J_pm^(x)}). From $J_- \otimes J_-^{*}$ the terms that contribute are
\begin{equation}
    J_x\otimes J_x + \frac{1}{4}(J_+^{(x)}\otimes J_+^{(x)}+J_-^{(x)}\otimes J_-^{(x)}),
\end{equation}
which can be rewritten using Eq.~(\ref{S2_defn}) as
\begin{equation}
    \frac{1}{4}\Bigl(2J_x\otimes J_x + J^2\otimes \mathbb{I}+\mathbb{I}\otimes J^2 - S^2\Bigr).
\end{equation}
Similarly, from $J_+J_-$ the terms that contribute can be shown to simplify to
\begin{equation}
    \frac{1}{2}(J_x^2+J^2)
\end{equation}
using Eq.~(\ref{J_pm^x expression with Jx}).
Combining these results yields an expression for the effective perturbation, $\mathcal{L}_D^{\mathrm{eff}}$, as
\begin{equation}
    \mathcal{L}_D^{\mathrm{eff}} = \frac{N\Gamma}{4}\Bigl(2J_x\otimes J_x - J_x^2 \otimes \mathbb{I} -\mathbb{I} \otimes J_x^2 - S^2\Bigr).
\end{equation}
Notice that $S_x^2=(J_x\otimes \mathbb{I}- \mathbb{I}\otimes J_x^T)^2= J_x^2 \otimes \mathbb{I}+ \mathbb{I}\otimes J_x^2 - 2J_x\otimes J_x$, where we used that $J_x^T=J_x$ if $J_x$ is diagonal. Using this, we obtain the required expression,
\begin{equation}
    \mathcal{L}_D^{\mathrm{eff}}= -\frac{N\Gamma}{4}\Bigl(S_x^2 +S^2\Bigr).
\end{equation}

\section{Evolution Equations for Spin Operators}
\label{appendix:evolution_eqn_spin_operators}

In this section, we derive the evolution equations for $\langle J_{\alpha}\rangle$. First, we focus on a general observable $\mathcal{O}$ and a general Lindblad master equation of the form
\begin{equation}
    \frac{d}{dt}\rho = -i [H_S, \rho]+\sum_i \gamma_i \Bigl(A_i \rho A_i^{\dagger} - \frac{1}{2}\{A_i^{\dagger}A_i, \rho \}\Bigr).
\end{equation}
We can obtain an equation purely in terms of expectation values using that $\frac{d}{dt}\langle \mathcal{O} \rangle  = \Tr(\mathcal{O} \frac{d}{dt}\rho)$. 
Assuming there is only a single Hermitian jump operator, and invoking the cyclic property of the trace allows us to write
\begin{equation}
     \frac{d}{dt}\langle\mathcal{O}\rangle = \Tr \Bigl\{\Bigl(
    i[H_S, \mathcal{O}]
    -  \frac{\gamma}{2} \bigl([A, A \mathcal{O}] + [\mathcal{O}A, A] 
    \bigr)
    \Bigr)\rho
    \Bigr\}.
\end{equation}
We demonstrate how to proceed with this by considering model B, such that now $\mathcal{O}=J_\alpha$. All the commutators can be evaluated, yielding the following set of coupled differential equations
\begin{equation}
\begin{split}
    & \frac{d}{dt}\langle J_x \rangle = 
    - 2 \kappa \langle J_x \rangle, \\
    & \frac{d}{dt}\langle J_y \rangle = 
   \omega \langle J_z \rangle - 2\kappa \langle J_y \rangle \\
    & \frac{d}{dt} \langle J_z \rangle, = 
    - \omega \langle J_y \rangle. \\
\end{split}
\end{equation}
$\langle J_x \rangle$ decouples from the other two components, which satisfy the following matrix equation
\begin{equation}
    \frac{d}{dt}\begin{pmatrix}
         \langle J_y\rangle \\
          \langle J_z \rangle \\
    \end{pmatrix}
    = 
    \underbrace{
    \begin{pmatrix}
        - 2\kappa &  \omega\\
        -\omega & 0 \\
    \end{pmatrix}
    }_{\mathbf{M}}
    \begin{pmatrix}
        \langle J_y\rangle \\
          \langle J_z \rangle \\
    \end{pmatrix}.
\end{equation}
The solution to this equation can be expressed by diagonalising $\mathbf{M}$, such that
\begin{equation}
    \begin{pmatrix}
        \langle J_y(t)\rangle \\
          \langle J_z(t) \rangle \\
    \end{pmatrix}
    = c_{+} e^{\lambda_+ t} \mathbf{v}_+ + c_{-} e^{\lambda_- t} \mathbf{v}_-,
\end{equation}
where $\lambda_{\pm}= - \kappa \pm i \omega\sqrt{1-(\kappa/\omega)^2}$ and $\mathbf{v}_{\pm}$ are the eigenvalues and eigenvectors of $\mathbf{M}$. Since $\mathbf{M}$ is a real matrix, this expression can be conveniently rewritten as 
\begin{equation}
    \begin{pmatrix}
        \langle J_y(t)\rangle \\
          \langle J_z(t) \rangle \\
    \end{pmatrix}
    = \mathrm{Re}\left(c\,  e^{\lambda_+t} 
    \begin{pmatrix}
        1 \\
        i \\
    \end{pmatrix}\right),
\end{equation}
where $c$ is a constant to be determined by the initial conditions. Thus, we see that both $\langle J_y(t)\rangle$ and $\langle J_z(t)\rangle$
exhibit decaying oscillations in the finite-$N$ limit, with a decay rate $\kappa$ and oscillation frequency $\omega\sqrt{1-(\kappa/\omega)^2}$. In addition, these two components are $\pi/2$ out of phase.

\newpage
\bibliography{references}

@article{Gyamfi_2020,
doi = {10.1088/1361-6404/ab9fdd},
url = {https://dx.doi.org/10.1088/1361-6404/ab9fdd},
year = {2020},
month = {oct},
publisher = {IOP Publishing},
volume = {41},
number = {6},
pages = {063002},
author = {Gyamfi, Jerryman A},
title = {Fundamentals of quantum mechanics in {Liouville} space},
journal = {European Journal of Physics}
}

@article{Perturbative_appraoch_open_quantum,
	author = {Li, Andy C. Y. and Petruccione, F. and Koch, Jens},
	date = {2014/05/08},
	doi = {10.1038/srep04887},
	id = {Li2014},
	isbn = {2045-2322},
	journal = {Scientific Reports},
	number = {1},
	pages = {4887},
	title = {{Perturbative} approach to {Markovian} open quantum systems},
	url = {https://doi.org/10.1038/srep04887},
	volume = {4},
	year = {2014}}

@book{Kato:1966:PTL,
  alias = {Kato 66},
  author = {Kato, Tosio},
  bibdate = {Fri Nov 24 15:18:30 1995},
  bibsource = {ftp://ftp.math.utah.edu/pub/bibnet/authors/m/matched-field-proc.bib},
  key = {Eigenvalues, Hilbert spaces},
  lccn = {QA320 .K33},
  pages = {74-81},
  sthbib = {M3 Kat 81 60},
  title = {Perturbation Theory for Linear Operators},
  year = 1966
}

@article{franco_nori_btc,
  title = {Symmetries and conserved quantities of boundary time crystals in generalized spin models},
  author = {Piccitto, Giulia and Wauters, Matteo and Nori, Franco and Shammah, Nathan},
  journal = {Phys. Rev. B},
  volume = {104},
  issue = {1},
  pages = {014307},
  numpages = {17},
  year = {2021},
  month = {Jul},
  publisher = {American Physical Society},
  doi = {10.1103/PhysRevB.104.014307},
  url = {https://link.aps.org/doi/10.1103/PhysRevB.104.014307}
}

@Article{SSB1,
	title={An introduction to spontaneous symmetry breaking},
	author={Aron J. Beekman and Louk Rademaker and Jasper van Wezel},
	journal={SciPost Phys. Lect. Notes},
	pages={11},
	year={2019},
	publisher={SciPost},
	doi={10.21468/SciPostPhysLectNotes.11},
	url={https://scipost.org/10.21468/SciPostPhysLectNotes.11},
}

@Article{SSB2,
AUTHOR = {Brauner, Tomáš},
TITLE = {{Spontaneous Symmetry Breaking} and {Nambu–Goldstone Bosons} in {Quantum Many-Body Systems}},
JOURNAL = {Symmetry},
VOLUME = {2},
YEAR = {2010},
NUMBER = {2},
PAGES = {609--657},
URL = {https://www.mdpi.com/2073-8994/2/2/609},
ISSN = {2073-8994},
DOI = {10.3390/sym2020609}
}

@article{SSB3,
title = {Spontaneous symmetry breaking in {Bose}–{Einstein} condensates},
journal = {Chemical Physics Letters},
volume = {288},
number = {2},
pages = {248-252},
year = {1998},
issn = {0009-2614},
doi = {https://doi.org/10.1016/S0009-2614(98)00270-X},
url = {https://www.sciencedirect.com/science/article/pii/S000926149800270X},
author = {Roger A. Hegstrom}
}

@article{wilczek,
  title = {{Quantum} {Time} {Crystals}},
  author = {Wilczek, Frank},
  journal = {Phys. Rev. Lett.},
  volume = {109},
  issue = {16},
  pages = {160401},
  numpages = {5},
  year = {2012},
  month = {Oct},
  publisher = {American Physical Society},
  doi = {10.1103/PhysRevLett.109.160401},
  url = {https://link.aps.org/doi/10.1103/PhysRevLett.109.160401}
}

@misc{khemani2019briefhistorytimecrystals,
      title={A {Brief} {History} of {Time} {Crystals}}, 
      author={Vedika Khemani and Roderich Moessner and S. L. Sondhi},
      year={2019},
      eprint={1910.10745},
      archivePrefix={arXiv},
      primaryClass={cond-mat.str-el},
      url={https://arxiv.org/abs/1910.10745}, 
}

@article{Sacha_2018,
doi = {10.1088/1361-6633/aa8b38},
url = {https://dx.doi.org/10.1088/1361-6633/aa8b38},
year = {2017},
month = {nov},
publisher = {IOP Publishing},
volume = {81},
number = {1},
pages = {016401},
author = {Sacha, Krzysztof and Zakrzewski, Jakub},
title = {Time crystals: a review},
journal = {Reports on Progress in Physics}
}

@article{JOHANSSON20121760,
title = {{QuTiP}: An open-source {Python} framework for the dynamics of open quantum systems},
journal = {Computer Physics Communications},
volume = {183},
number = {8},
pages = {1760-1772},
year = {2012},
issn = {0010-4655},
doi = {https://doi.org/10.1016/j.cpc.2012.02.021},
url = {https://www.sciencedirect.com/science/article/pii/S0010465512000835},
author = {J.R. Johansson and P.D. Nation and Franco Nori}
}

@article{JOHANSSON20131234,
title = {{QuTiP} 2: A {Python} framework for the dynamics of open quantum systems},
journal = {Computer Physics Communications},
volume = {184},
number = {4},
pages = {1234-1240},
year = {2013},
issn = {0010-4655},
doi = {https://doi.org/10.1016/j.cpc.2012.11.019},
url = {https://www.sciencedirect.com/science/article/pii/S0010465512003955},
author = {J.R. Johansson and P.D. Nation and Franco Nori}
}

@misc{lambert2024qutip5quantumtoolbox,
      title={{QuTiP} 5: The {Quantum Toolbox} in {Python}}, 
      author={Neill Lambert and Eric Giguère and Paul Menczel and Boxi Li and Patrick Hopf and Gerardo Suárez and Marc Gali and Jake Lishman and Rushiraj Gadhvi and Rochisha Agarwal and Asier Galicia and Nathan Shammah and Paul Nation and J. R. Johansson and Shahnawaz Ahmed and Simon Cross and Alexander Pitchford and Franco Nori},
      year={2024},
      eprint={2412.04705},
      archivePrefix={arXiv},
      primaryClass={quant-ph},
      url={https://arxiv.org/abs/2412.04705}, 
}

@article{significant_debate_1,
  title = {{Space-Time Crystals} of {Trapped Ions}},
  author = {Li, Tongcang and Gong, Zhe-Xuan and Yin, Zhang-Qi and Quan, H. T. and Yin, Xiaobo and Zhang, Peng and Duan, L.-M. and Zhang, Xiang},
  journal = {Phys. Rev. Lett.},
  volume = {109},
  issue = {16},
  pages = {163001},
  numpages = {5},
  year = {2012},
  month = {Oct},
  publisher = {American Physical Society},
  doi = {10.1103/PhysRevLett.109.163001},
  url = {https://link.aps.org/doi/10.1103/PhysRevLett.109.163001}
}

@article{significant_debate_2,
  title = {Comment on ``{Space-Time Crystals} of {Trapped Ions}''},
  author = {Bruno, Patrick},
  journal = {Phys. Rev. Lett.},
  volume = {111},
  issue = {2},
  pages = {029301},
  numpages = {1},
  year = {2013},
  month = {Jul},
  publisher = {American Physical Society},
  doi = {10.1103/PhysRevLett.111.029301},
  url = {https://link.aps.org/doi/10.1103/PhysRevLett.111.029301}
}

@article{significant_debate_3,
doi = {10.1209/0295-5075/103/57008},
url = {https://dx.doi.org/10.1209/0295-5075/103/57008},
year = {2013},
month = {sep},
publisher = {EDP Sciences, IOP Publishing and Società Italiana di Fisica},
volume = {103},
number = {5},
pages = {57008},
author = {Nozières, Philippe},
title = {Time crystals: Can diamagnetic currents drive a charge density wave into rotation?},
journal = {Europhysics Letters}
}

@article{significant_debate_4,
	author = {Volovik, G.  E. },
	date = {2013/12/01},
	doi = {10.1134/S0021364013210133},
	id = {Volovik2013},
	isbn = {1090-6487},
	journal = {JETP Letters},
	number = {8},
	pages = {491--495},
	title = {On the broken time translation symmetry in macroscopic systems: Precessing states and off-diagonal long-range order},
	url = {https://doi.org/10.1134/S0021364013210133},
	volume = {98},
	year = {2013}}

@article{no_go_theorem,
  title = {{Absence} of {Quantum Time Crystals}},
  author = {Watanabe, Haruki and Oshikawa, Masaki},
  journal = {Phys. Rev. Lett.},
  volume = {114},
  issue = {25},
  pages = {251603},
  numpages = {5},
  year = {2015},
  month = {Jun},
  publisher = {American Physical Society},
  doi = {10.1103/PhysRevLett.114.251603},
  url = {https://link.aps.org/doi/10.1103/PhysRevLett.114.251603}
}

@article{Long_range_time_crystals,
  title = {{Quantum Time Crystals} from {Hamiltonians} with {Long-Range Interactions}},
  author = {Kozin, Valerii K. and Kyriienko, Oleksandr},
  journal = {Phys. Rev. Lett.},
  volume = {123},
  issue = {21},
  pages = {210602},
  numpages = {6},
  year = {2019},
  month = {Nov},
  publisher = {American Physical Society},
  doi = {10.1103/PhysRevLett.123.210602},
  url = {https://link.aps.org/doi/10.1103/PhysRevLett.123.210602}
}

@article{DTC1,
  title = {Modeling spontaneous breaking of time-translation symmetry},
  author = {Sacha, Krzysztof},
  journal = {Phys. Rev. A},
  volume = {91},
  issue = {3},
  pages = {033617},
  numpages = {5},
  year = {2015},
  month = {Mar},
  publisher = {American Physical Society},
  doi = {10.1103/PhysRevA.91.033617},
  url = {https://link.aps.org/doi/10.1103/PhysRevA.91.033617}
}

@article{DTC2,
  title = {{Floquet Time Crystals}},
  author = {Else, Dominic V. and Bauer, Bela and Nayak, Chetan},
  journal = {Phys. Rev. Lett.},
  volume = {117},
  issue = {9},
  pages = {090402},
  numpages = {5},
  year = {2016},
  month = {Aug},
  publisher = {American Physical Society},
  doi = {10.1103/PhysRevLett.117.090402},
  url = {https://link.aps.org/doi/10.1103/PhysRevLett.117.090402}
}

@article{DTC3,
  title = {{Phase Structure} of {Driven Quantum Systems}},
  author = {Khemani, Vedika and Lazarides, Achilleas and Moessner, Roderich and Sondhi, S. L.},
  journal = {Phys. Rev. Lett.},
  volume = {116},
  issue = {25},
  pages = {250401},
  numpages = {6},
  year = {2016},
  month = {Jun},
  publisher = {American Physical Society},
  doi = {10.1103/PhysRevLett.116.250401},
  url = {https://link.aps.org/doi/10.1103/PhysRevLett.116.250401}
}

@article{DTC4,
  title = {Absolute stability and spatiotemporal long-range order in {Floquet} systems},
  author = {von Keyserlingk, C. W. and Khemani, Vedika and Sondhi, S. L.},
  journal = {Phys. Rev. B},
  volume = {94},
  issue = {8},
  pages = {085112},
  numpages = {11},
  year = {2016},
  month = {Aug},
  publisher = {American Physical Society},
  doi = {10.1103/PhysRevB.94.085112},
  url = {https://link.aps.org/doi/10.1103/PhysRevB.94.085112}
}

@article{DTC5,
  title = {Defining time crystals via representation theory},
  author = {Khemani, Vedika and von Keyserlingk, C. W. and Sondhi, S. L.},
  journal = {Phys. Rev. B},
  volume = {96},
  issue = {11},
  pages = {115127},
  numpages = {7},
  year = {2017},
  month = {Sep},
  publisher = {American Physical Society},
  doi = {10.1103/PhysRevB.96.115127},
  url = {https://link.aps.org/doi/10.1103/PhysRevB.96.115127}
}

@article{DTC6,
  title = {{Prethermal Phases} of {Matter Protected} by {Time-Translation Symmetry}},
  author = {Else, Dominic V. and Bauer, Bela and Nayak, Chetan},
  journal = {Phys. Rev. X},
  volume = {7},
  issue = {1},
  pages = {011026},
  numpages = {21},
  year = {2017},
  month = {Mar},
  publisher = {American Physical Society},
  doi = {10.1103/PhysRevX.7.011026},
  url = {https://link.aps.org/doi/10.1103/PhysRevX.7.011026}
}

@article{DTC7,
  title = {{Discrete Time Crystals: Rigidity, Criticality}, and {Realizations}},
  author = {Yao, N. Y. and Potter, A. C. and Potirniche, I.-D. and Vishwanath, A.},
  journal = {Phys. Rev. Lett.},
  volume = {118},
  issue = {3},
  pages = {030401},
  numpages = {6},
  year = {2017},
  month = {Jan},
  publisher = {American Physical Society},
  doi = {10.1103/PhysRevLett.118.030401},
  url = {https://link.aps.org/doi/10.1103/PhysRevLett.118.030401}
}

@article{DTC8,
  title = {{Clean Floquet Time Crystals}: {Models} and {Realizations} in {Cold Atoms}},
  author = {Huang, Biao and Wu, Ying-Hai and Liu, W. Vincent},
  journal = {Phys. Rev. Lett.},
  volume = {120},
  issue = {11},
  pages = {110603},
  numpages = {7},
  year = {2018},
  month = {Mar},
  publisher = {American Physical Society},
  doi = {10.1103/PhysRevLett.120.110603},
  url = {https://link.aps.org/doi/10.1103/PhysRevLett.120.110603}
}

@article{DTC9,
  title = {Floquet time crystal in the {Lipkin-Meshkov-Glick} model},
  author = {Russomanno, Angelo and Iemini, Fernando and Dalmonte, Marcello and Fazio, Rosario},
  journal = {Phys. Rev. B},
  volume = {95},
  issue = {21},
  pages = {214307},
  numpages = {12},
  year = {2017},
  month = {Jun},
  publisher = {American Physical Society},
  doi = {10.1103/PhysRevB.95.214307},
  url = {https://link.aps.org/doi/10.1103/PhysRevB.95.214307}
}

@article{DTC10,
	author = {Else, Dominic V. and Monroe, Christopher and Nayak, Chetan and Yao, Norman Y.},
	doi = {https://doi.org/10.1146/annurev-conmatphys-031119-050658},
	issn = {1947-5462},
	journal = {Annual Review of Condensed Matter Physics},
	number = {Volume 11, 2020},
	pages = {467-499},
	publisher = {Annual Reviews},
	title = {{Discrete Time Crystals}},
	type = {Journal Article},
	url = {https://www.annualreviews.org/content/journals/10.1146/annurev-conmatphys-031119-050658},
	volume = {11},
	year = {2020}}

@article{DTC_EXPT_1,
	author = {Zhang, J. and Hess, P. W. and Kyprianidis, A. and Becker, P. and Lee, A. and Smith, J. and Pagano, G. and Potirniche, I. -D. and Potter, A. C. and Vishwanath, A. and Yao, N. Y. and Monroe, C.},
	date = {2017/03/01},
	doi = {10.1038/nature21413},
	id = {Zhang2017},
	isbn = {1476-4687},
	journal = {Nature},
	number = {7644},
	pages = {217--220},
	title = {Observation of a discrete time crystal},
	url = {https://doi.org/10.1038/nature21413},
	volume = {543},
	year = {2017}}

@article{DTC_EXPT_2,
	author = {Choi, Soonwon and Choi, Joonhee and Landig, Renate and Kucsko, Georg and Zhou, Hengyun and Isoya, Junichi and Jelezko, Fedor and Onoda, Shinobu and Sumiya, Hitoshi and Khemani, Vedika and von Keyserlingk, Curt and Yao, Norman Y. and Demler, Eugene and Lukin, Mikhail D.},
	date = {2017/03/01},
	doi = {10.1038/nature21426},
	id = {Choi2017},
	isbn = {1476-4687},
	journal = {Nature},
	number = {7644},
	pages = {221--225},
	title = {Observation of discrete time-crystalline order in a disordered dipolar many-body system},
	url = {https://doi.org/10.1038/nature21426},
	volume = {543},
	year = {2017}}

@article{DTC_EXPT_3,
	author = {Taheri, Hossein and Matsko, Andrey B. and Maleki, Lute and Sacha, Krzysztof},
	date = {2022/02/14},
	doi = {10.1038/s41467-022-28462-x},
	id = {Taheri2022},
	isbn = {2041-1723},
	journal = {Nature Communications},
	number = {1},
	pages = {848},
	title = {All-optical dissipative discrete time crystals},
	url = {https://doi.org/10.1038/s41467-022-28462-x},
	volume = {13},
	year = {2022}}

@article{DTC_EXPT_4,
	author = {A. Kyprianidis and F. Machado and W. Morong and P. Becker and K. S. Collins and D. V. Else and L. Feng and P. W. Hess and C. Nayak and G. Pagano and N. Y. Yao and C. Monroe},
	doi = {10.1126/science.abg8102},
	journal = {Science},
	number = {6547},
	pages = {1192-1196},
	title = {Observation of a prethermal discrete time crystal},
	volume = {372},
	year = {2021},
}

@article{DTC_EXPT_5,
	author = {Philipp Frey and Stephan Rachel},
	doi = {10.1126},
	journal = {Science Advances},
	number = {9},
	title = {Realization of a discrete time crystal on 57 qubits of a quantum computer},
	volume = {8},
	year = {2022},
}

@article{BTC1,
  title = {{Boundary Time Crystals}},
  author = {Iemini, F. and Russomanno, A. and Keeling, J. and Schir\`o, M. and Dalmonte, M. and Fazio, R.},
  journal = {Phys. Rev. Lett.},
  volume = {121},
  issue = {3},
  pages = {035301},
  numpages = {6},
  year = {2018},
  month = {Jul},
  publisher = {American Physical Society},
  doi = {10.1103/PhysRevLett.121.035301},
  url = {https://link.aps.org/doi/10.1103/PhysRevLett.121.035301}
}

@article{BTCs_d_level_system,
  title = {Boundary time crystals in collective $d$-level systems},
  author = {Prazeres, Luis Fernando dos and Souza, Leonardo da Silva and Iemini, Fernando},
  journal = {Phys. Rev. B},
  volume = {103},
  issue = {18},
  pages = {184308},
  numpages = {16},
  year = {2021},
  month = {May},
  publisher = {American Physical Society},
  doi = {10.1103/PhysRevB.103.184308},
  url = {https://link.aps.org/doi/10.1103/PhysRevB.103.184308}
}

@book{breuer_open_qm,
	author = {Breuer, Heinz-Peter and Petruccione, Francesco},
	doi = {10.1093/acprof:oso/9780199213900.001.0001},
	isbn = {9780199213900},
	month = {01},
	publisher = {Oxford University Press},
	title = {The Theory of Open Quantum Systems},
	url = {https://doi.org/10.1093/acprof:oso/9780199213900.001.0001},
	year = {2007}}

@article{BTC_semiclassical_1,
  title = {Emergent limit cycles and time crystal dynamics in an atom-cavity system},
  author = {Ke\ss{}ler, Hans and Cosme, Jayson G. and Hemmerling, Michal and Mathey, Ludwig and Hemmerich, Andreas},
  journal = {Phys. Rev. A},
  volume = {99},
  issue = {5},
  pages = {053605},
  numpages = {6},
  year = {2019},
  month = {May},
  publisher = {American Physical Society},
  doi = {10.1103/PhysRevA.99.053605},
  url = {https://link.aps.org/doi/10.1103/PhysRevA.99.053605}
}

@article{BTC_quantum_trajectories,
  title = {Revealing the nature of nonequilibrium phase transitions with quantum trajectories},
  author = {Link, Valentin and Luoma, Kimmo and Strunz, Walter T.},
  journal = {Phys. Rev. A},
  volume = {99},
  issue = {6},
  pages = {062120},
  numpages = {7},
  year = {2019},
  month = {Jun},
  publisher = {American Physical Society},
  doi = {10.1103/PhysRevA.99.062120},
  url = {https://link.aps.org/doi/10.1103/PhysRevA.99.062120}
}

@article{exact_differential_method,
  title = {{Integrable Quantum Dynamics} of {Open Collective Spin Models}},
  author = {Ribeiro, Pedro and Prosen, Toma\ifmmode \check{z}\else \v{z}\fi{}},
  journal = {Phys. Rev. Lett.},
  volume = {122},
  issue = {1},
  pages = {010401},
  numpages = {6},
  year = {2019},
  month = {Jan},
  publisher = {American Physical Society},
  doi = {10.1103/PhysRevLett.122.010401},
  url = {https://link.aps.org/doi/10.1103/PhysRevLett.122.010401}
}

@article{btc_found,
  title = {Exceptional spectral phase in a dissipative collective spin model},
  author = {Rubio-Garc\'{\i}a, \'Alvaro and Corps, \'Angel L. and Rela\~no, Armando and Molina, Rafael A. and P\'erez-Bernal, Francisco and Garc\'{\i}a-Ramos, Jos\'e Enrique and Dukelsky, Jorge},
  journal = {Phys. Rev. A},
  volume = {106},
  issue = {1},
  pages = {L010201},
  numpages = {6},
  year = {2022},
  month = {Jul},
  publisher = {American Physical Society},
  doi = {10.1103/PhysRevA.106.L010201},
  url = {https://link.aps.org/doi/10.1103/PhysRevA.106.L010201}
}

@article{btc_numerical_1,
  title = {Dissipative time crystal in an asymmetric nonlinear photonic dimer},
  author = {Seibold, Kilian and Rota, Riccardo and Savona, Vincenzo},
  journal = {Phys. Rev. A},
  volume = {101},
  issue = {3},
  pages = {033839},
  numpages = {9},
  year = {2020},
  month = {Mar},
  publisher = {American Physical Society},
  doi = {10.1103/PhysRevA.101.033839},
  url = {https://link.aps.org/doi/10.1103/PhysRevA.101.033839}
}

@article{btc_numerical_2,
doi = {10.1088/1367-2630/ab9ae3},
url = {https://dx.doi.org/10.1088/1367-2630/ab9ae3},
year = {2020},
month = {jul},
publisher = {IOP Publishing},
volume = {22},
number = {7},
pages = {075002},
author = {Lledó, Cristóbal and Szymańska, Marzena H},
title = {A dissipative time crystal with or without {Z2} symmetry breaking},
journal = {New Journal of Physics}
}

@article{btc_numerical_3,
  title = {Genuine multipartite correlations in a boundary time crystal},
  author = {Louren\ifmmode \mbox{\c{c}}\else \c{c}\fi{}o, Ant\^onio C. and Prazeres, Luis Fernando dos and Maciel, Thiago O. and Iemini, Fernando and Duzzioni, Eduardo I.},
  journal = {Phys. Rev. B},
  volume = {105},
  issue = {13},
  pages = {134422},
  numpages = {13},
  year = {2022},
  month = {Apr},
  publisher = {American Physical Society},
  doi = {10.1103/PhysRevB.105.134422},
  url = {https://link.aps.org/doi/10.1103/PhysRevB.105.134422}
}

@article{btc_semiclassical_2,
  title = {Exact solution of a boundary time-crystal phase transition: Time-translation symmetry breaking and non-Markovian dynamics of correlations},
  author = {Carollo, Federico and Lesanovsky, Igor},
  journal = {Phys. Rev. A},
  volume = {105},
  issue = {4},
  pages = {L040202},
  numpages = {6},
  year = {2022},
  month = {Apr},
  publisher = {American Physical Society},
  doi = {10.1103/PhysRevA.105.L040202},
  url = {https://link.aps.org/doi/10.1103/PhysRevA.105.L040202}
}

@article{dissipative_dynamics_xxz_rg,
  title = {Dissipative dynamics in open {XXZ Richardson-Gaudin} models},
  author = {Claeys, Pieter W. and Lamacraft, Austen},
  journal = {Phys. Rev. Res.},
  volume = {4},
  issue = {1},
  pages = {013033},
  numpages = {11},
  year = {2022},
  month = {Jan},
  publisher = {American Physical Society},
  doi = {10.1103/PhysRevResearch.4.013033},
  url = {https://link.aps.org/doi/10.1103/PhysRevResearch.4.013033}
}

@article{btc_semiclassical_3,
  title = {{Dissipative} time crystals with long-range {Lindbladians}},
  author = {Passarelli, Gianluca and Lucignano, Procolo and Fazio, Rosario and Russomanno, Angelo},
  journal = {Phys. Rev. B},
  volume = {106},
  issue = {22},
  pages = {224308},
  numpages = {13},
  year = {2022},
  month = {Dec},
  publisher = {American Physical Society},
  doi = {10.1103/PhysRevB.106.224308},
  url = {https://link.aps.org/doi/10.1103/PhysRevB.106.224308}
}

@article{gapless_excitations_cumulant_expansion,
  title = {{Sufficient Condition} for {Gapless Spin-Boson Lindbladians}, and {Its Connection to Dissipative Time Crystals}},
  author = {Souza, Leonardo da Silva and dos Prazeres, Luis Fernando and Iemini, Fernando},
  journal = {Phys. Rev. Lett.},
  volume = {130},
  issue = {18},
  pages = {180401},
  numpages = {7},
  year = {2023},
  month = {May},
  publisher = {American Physical Society},
  doi = {10.1103/PhysRevLett.130.180401},
  url = {https://link.aps.org/doi/10.1103/PhysRevLett.130.180401}
}

@article{btc_briefly_mentioned,
  title = {Dynamics of inhomogeneous spin ensembles with all-to-all interactions: Breaking permutational invariance},
  author = {Iemini, Fernando and Chang, Darrick and Marino, Jamir},
  journal = {Phys. Rev. A},
  volume = {109},
  issue = {3},
  pages = {032204},
  numpages = {11},
  year = {2024},
  month = {Mar},
  publisher = {American Physical Society},
  doi = {10.1103/PhysRevA.109.032204},
  url = {https://link.aps.org/doi/10.1103/PhysRevA.109.032204}
}

@misc{wang2025boundarytimecrystalsinduced,
      title={{Boundary Time Crystals Induced} by {Local Dissipation} and {Long-Range Interactions}}, 
      author={Zhuqing Wang and Ruochen Gao and Xiaoling Wu and Berislav Buča and Klaus Mølmer and Li You and Fan Yang},
      year={2025},
      eprint={2503.20761},
      archivePrefix={arXiv},
      primaryClass={quant-ph},
      url={https://arxiv.org/abs/2503.20761}, 
}

@article{dissipative_freezing,
  title = {Symmetries and conservation laws in quantum trajectories: Dissipative freezing},
  author = {S\'anchez Mu\~noz, Carlos and Bu\ifmmode \check{c}\else \v{c}\fi{}a, Berislav and Tindall, Joseph and Gonz\'alez-Tudela, Alejandro and Jaksch, Dieter and Porras, Diego},
  journal = {Phys. Rev. A},
  volume = {100},
  issue = {4},
  pages = {042113},
  numpages = {19},
  year = {2019},
  month = {Oct},
  publisher = {American Physical Society},
  doi = {10.1103/PhysRevA.100.042113},
  url = {https://link.aps.org/doi/10.1103/PhysRevA.100.042113}
}

@article{btc_quantum_thermo,
doi = {10.1088/2058-9565/ad3f42},
url = {https://dx.doi.org/10.1088/2058-9565/ad3f42},
year = {2024},
month = {may},
publisher = {IOP Publishing},
volume = {9},
number = {3},
pages = {035024},
author = {Carollo, Federico and Lesanovsky, Igor and Antezza, Mauro and De Chiara, Gabriele},
title = {Quantum thermodynamics of boundary time-crystals},
journal = {Quantum Science and Technology}
}

@article{btcs_pt_nakanishi,
  title = {Dissipative time crystals originating from parity-time symmetry},
  author = {Nakanishi, Yuma and Sasamoto, Tomohiro},
  journal = {Phys. Rev. A},
  volume = {107},
  issue = {1},
  pages = {L010201},
  numpages = {7},
  year = {2023},
  month = {Jan},
  publisher = {American Physical Society},
  doi = {10.1103/PhysRevA.107.L010201},
  url = {https://link.aps.org/doi/10.1103/PhysRevA.107.L010201}
}

@article{Booker_2020,
doi = {10.1088/1367-2630/ababc4},
url = {https://doi.org/10.1088/1367-2630/ababc4},
year = {2020},
month = {aug},
publisher = {IOP Publishing},
volume = {22},
number = {8},
pages = {085007},
author = {Booker, Cameron and Buča, Berislav and Jaksch, Dieter},
title = {Non-stationarity and dissipative time crystals: spectral properties and finite-size effects},
journal = {New Journal of Physics}
}

@misc{nemeth2025btc-code,
  author       = {Dominik Nemeth},
 title        = {Python code: Analytical solution of boundary time crystals via the superspin basis},
  year         = {2025},
  howpublished = {\url{https://github.com/d-nemeth/boundary-time-crystals}}
}

\end{document}